\documentstyle[preprint,tighten,eqsecnum,prc,aps]{revtex}
\begin{document}
\hyphenation{Nijmegen}
\hyphenation{Rijken}
\draft
\preprint{\vbox{. \hfill ADP-98-37-T310}}

\title{Soft-core hyperon-nucleon potentials}
\author{Th.A.\ Rijken}
\address{Institute for Theoretical Physics, University of Nijmegen,
         Nijmegen, The Netherlands}
\author{V.G.J.\ Stoks}
\address{Physics Division, Argonne National Laboratory, Argonne,
         Illinois 60439 \protect\\ and
         Centre for the Subatomic Structure of Matter,
         University of Adelaide, SA 5005, Australia}
\author{Y.\ Yamamoto}
\address{Physics Section, Tsuru University, Tsuru, Yamanashi 402-0054, Japan}
\date{}
\maketitle

\begin{abstract}
A new Nijmegen soft-core OBE potential model is presented for the
low-energy $Y\!N$ interactions. Besides the results for the fit to
the scattering data, which largely defines the model, we also present
some applications to hypernuclear systems using the G-matrix method.
The potentials are generated by the exchange of nonets of pseudoscalar,
vector, and scalar mesons. As standard in the Nijmegen soft-core
models, we also include the $J=0$ contributions from the tensor
$f_{2},f'_{2},a_{2}$ and pomeron Regge trajectories, and use Gaussian
form factors to guarantee that the potentials have a soft behavior near
the origin.
An important innovation with respect to the original soft-core potential
is the assignment of the cut-off masses for the baryon-baryon-meson
(BBM) vertices in accordance with broken SU(3)$_F$, which serves to
connect the $N\!N$ and the $Y\!N$ channels.
As a novel feature, we allow for medium strong breaking of the coupling
constants, using the $^{3}P_{0}$ model with a Gell-Mann--Okubo
hypercharge breaking for the BBM coupling. Charge-symmetry breaking in
the $\Lambda p$ and $\Lambda n$ channels is included as well.
We present six hyperon-nucleon potentials which describe the
available $Y\!N$ cross section data equally well, but which exhibit
some differences on a more detailed level. The differences are
constructed such that the models encompass a range of scattering
lengths in the $\Sigma N$ and $\Lambda N$ channels.
In all cases, we obtained $\chi^{2}/N_{\rm data}\approx0.55$ for 35 $Y\!N$
data. In particular, we were able to fit the precise experimental datum
$r_{R}=0.468\pm0.010$ for the inelastic capture ratio at rest.
For the scalar-meson mixing angle we obtained values
$\theta_S=37^\circ$--$40^\circ$, which points to almost ideal
mixing angles for the scalar $q\bar{q}$ states.
The G-matrix results indicate that the remarkably different
spin-spin terms of the six potentials appear specifically
in the energy spectra of $\Lambda$ hypernuclei.
\end{abstract}

\narrowtext

\section{Introduction}
In Refs.~\cite{Nag78,Mae89}, henceforth referred to as I and II,
respectively, it has been shown that a soft-core one-boson-exchange
(OBE) model, based on Regge-pole theory~\cite{Rij85}, provides an
excellent simultaneous description of the rich and accurate
nucleon-nucleon ($N\!N$) and the more scarce hyperon-nucleon ($Y\!N$)
low-energy scattering data. However, in the application to the
hypernuclear systems using the G-matrix method, it was found that
the spin-spin interaction in the $\Lambda N$ channels needs a
correction~\cite{Yam90,Mot95,Has95}.
Another inconvenience with I and II is that an extension to the
$\Lambda\Lambda$ and $\Xi N$ channels cannot be done without the
introduction of extra free parameters.

In order to improve the soft-core interaction on these points, we here
modify the original soft-core OBE models of I and II in the following
way. First, we assign the cut-off parameters in the form factors for the
individual baryon-baryon-meson (BBM) vertices, constrained by broken
SU(3)$_F$ symmetry. This in contrast to I and II, where these cut-off
parameters were assigned per baryon-baryon SU(3)$_F$-irrep. Because
the $\Lambda\Lambda$ and $\Xi N$ channels involve the $\{1\}$-irrep,
which does not occur in the $N\!N$ and $Y\!N$ channels, the description
of these channels would require the introduction of additional free
parameters. However, there are no experimental scattering data to
determine these parameters. (The only experimental information on the
$\Lambda\Lambda$ interaction is limited to the ground states of
double-$\Lambda$ hypernuclei, but such information is ``contaminated''
by few-body effects.)
Second, we note that in Ref.~\cite{Mae89} the magnetic $F/(F+D)$ ratio
$\alpha_V^m$ for the vector mesons was fixed to its SU(6) value.
Therefore, in order to improve the spin-spin interaction, we here consider
$\alpha_V^m$ as a free input and make fits for different values
of this parameter. It turns out that this allows us to construct $Y\!N$
models which encompass a range of scattering lengths in the $^{1}S_{0}$
and the $^{3}S_{1}$ $\Lambda N$ channels. It is found that various
other quantities, calculated with these new models, also exhibit an
impressive correlation with the choice for $\alpha_V^m$.
By testing these models in hypernuclear systems we can select the
successful spin-spin interaction.
In order to have enough flexibility, we introduced a third modification
with respect to I and II; namely, we allow for medium strong breaking of
the coupling constants. The breaking is implemented according to the
$^{3}P_{0}$ model~\cite{Mic69} with a Gell-Mann--Okubo hypercharge
breaking.

Apart from the modifications indicated above, the OBE models of this
paper, henceforth referred to as NSC97 models, are motivated according
to the same physical principles as those of I and II.
We refer to the latter papers~\cite{Nag78,Mae89} for a more detailed
description of the physics background of the Nijmegen soft-core
baryon-baryon models. We here only briefly reiterate the main points.

The baryon-baryon soft-core models can be fully derived in the context
of the analytical $S$-matrix theory~\cite{Rij85}. This seems a proper
framework to describe baryons and mesons, which are composite particles.
In particular, in QCD the mesons are $q\bar{q}$ systems and any reasonable
interaction used in a Bethe-Salpeter approach to the $q\bar{q}$ systems
leads to mesons on Regge trajectories.
The consequences of the Regge trajectories for low-energy scattering
and the corresponding (relativistic) Lippmann-Schwinger equations
can be worked out in a consistent manner in the mentioned framework.

With a combined treatment of the $N\!N$ and $Y\!N$ channels we aim at
a high-quality description of the baryon-baryon interactions.
By high quality we mean a fit to the $Y\!N$ scattering data with a low
$\chi^{2}$, such that, while keeping the constraints forced on the
potentials by the fit to the $N\!N$ scattering data, the free parameters
with a clear physical significance (like, e.g., the $F/(F+D)$ ratios
$\alpha_{PV}$ and $\alpha_V^m$) assume realistic values.
Such a combined study of all baryon-baryon interactions, and especially
$N\!N$ and $Y\!N$, is desirable if one wants to test the assumption of
SU(3)$_F$ symmetry. For example, we want to investigate the properties
of the scalar mesons [$\varepsilon(760)$, $f_{0}(975)$, $a_{0}(980)$,
$\kappa(880)$], since especially the status of the scalar nonet is at
present not established yet. We also want to extract information about
scattering lengths, effective ranges, and the existence of resonances.
This, in spite of the scarce experimental $Y\!N$ data.
Moreover, we aim to extend the theoretical description to the
$\Lambda\Lambda$ and $\Xi N$ channels, where experiments may be
realized in the foreseeable future.

In this paper we treat in detail the following $Y\!N$ reactions for
which experimental data exist: (i) The coupled-channel reaction
$\Lambda p \Rightarrow \Lambda p, \Sigma^{+} n, \Sigma^{0} p$, below
the threshold of the coupling to the $\Sigma N$ channels;
(ii) The coupled-channel reaction $\Sigma^{-}p \Rightarrow \Lambda n,
\Sigma^{0}n, \Sigma^{-}p$, and (iii) The single-channel reaction
$\Sigma^{+}p \Rightarrow \Sigma^{+}p$.
The NSC97 models of this paper are a step forward in the realization
of a program where the baryon-baryon interactions for scattering and
hypernuclei can be described in the context of broken SU(3)$_F$ symmetry.

For definiteness, we list the meson exchanges which are included:

(i) The pseudoscalar mesons ($\pi$, $\eta$, $\eta'$, $K$), with the
$\eta$--$\eta'$ mixing angle $\theta_{PV}=-23.0^{0}$ from the
Gell-Mann--Okubo mass formula. The $F/(F+D)$ ratio, $\alpha_{PV}=0.355$,
is given by the value found in semileptonic weak decays~\cite{Dum83}.

(ii) The vector mesons ($\rho$, $\phi$, $\omega$, $K^{\star}$), with
the $\phi$--$\omega$ mixing angle $\theta_V=37.5^\circ$~\cite{Dum83}
and the electric $\alpha_V^e=1$, which follows the ``universality''
assumption~\cite{Sak60}. The magnetic $\alpha_V^m$ is used
as a free input to encompass a range of scattering lengths,
characterizing the different models, but is restricted to values
consistent with static or relativistic SU(6) predictions~\cite{Sak65}.

(iii) The scalar mesons [$a_{0}(980)$, $f_{0}(975)$, $f_{0}(760)$,
$\kappa(880)$]. In the following, we will reserve $f_0$ for the
$f_0(975)$ meson and use $\varepsilon$ for the $f_0(760)$ meson.
The free $f_{0}$--$\varepsilon$ mixing angle $\theta_S$ is to be
determined in the fit to the $Y\!N$ data.

(iv) The ``diffractive'' contribution from the pomeron $P$ and from the
tensor mesons [$f_{2}(1285)$, $f'_{2}(1525)$, $a_{2}(1270)$]. These
exchanges will give repulsive contributions of a Gaussian type.

The BBM vertices are described by coupling constants and form factors,
which correspond to the Regge residues at high energies~\cite{Rij85}.
The form factors are taken to be of the Gaussian type, like the residue
functions in many Regge-pole models for high-energy scattering.
Note that also in (nonrelativistic) quark models a Gaussian behavior
of the form factors is most natural. These form factors evidently
guarantee a soft behavior of the potentials in configuration
space at small distances.

It turns out that, starting from the soft-core OBE model for the $N\!N$
interaction, we are indeed able to achieve a very good description of
the $Y\!N$ data, and at the same time maintain values for the free
parameters which are consistent with the present view on low-energy
hadron physics.
Like in I and II we use SU(3)$_F$ symmetry for the coupling constants,
while SU(3)$_F$ breaking is included by (i) using the physical masses
of the mesons and baryons in the potentials and in the Schr\"odinger
equation; (ii) allowing for meson-mixing within a nonet ($\eta$--$\eta'$,
$\omega$--$\phi$, $\varepsilon$--$f_{0}$); (iii) including charge-symmetry
breaking (CSB) due to $\Lambda$-$\Sigma^0$ mixing~\cite{Dal64}, which
introduces a one-pion-exchange (OPE) potential in the $\Lambda p$ and
$\Lambda n$ channels; and (iv) taking into account the Coulomb interaction.
In order to include the Coulomb interaction exactly, and to account
as much as possible for the mass differences between the baryons, we
solve the multichannel Schr\"odinger equation on the physical particle
basis. However, in order to limit the number of different form factors,
the nuclear potentials are calculated on the isospin basis.
This means that we include only the so-called ``medium strong''
SU(3)$_F$ breaking in the potentials.

The content of this paper is as follows. In Sec.~\ref{sec:def} we
give the meson-baryon interaction Lagrangian and define the OBE
potentials for the Lippmann-Schwinger equation.
In Sec.~\ref{sec:YNchan} we review the possible $Y\!N$ channels that
are allowed and discuss some aspects of the multichannel Schr\"odinger
equation. In Sec.~\ref{sec:mesons} we discuss the pseudoscalar- and
vector-meson multiplets. The scalar-meson multiplet is discussed rather
extensively, because of its important role in the soft-core OBE models.
Also, some remarks are made on the origin and nature of the pomeron and
tensor-meson contributions.
In Sec.~\ref{sec:su3break} we outline the broken SU(3)$_F$ scheme of
the form factors and the coupling constants, in particular the employed
$^3P_0$ model.
Section~\ref{sec:YNfit} contains the results of the fits to the $Y\!N$
scattering data, while in Sec.~\ref{sec:hypnuc} the properties of the
models are investigated in hypernuclear systems within the G-matrix
approach. Finally, in Sec.~\ref{sec:conc} we finish with a final
discussion and draw some conclusions.

\section{Definition of the potentials}
\label{sec:def}
The nucleon-nucleon ($N\!N$) and hyperon-nucleon ($Y\!N$) potentials
constitute only a subset of possible interaction channels for the
baryon-baryon interaction; they cover only the strangeness $S=0,-1$
channels. The various members of the baryon octet, in principle, allow
for baryon-baryon interactions with total strangeness up to $S=-4$.
Since at present there are no scattering data for the $S=-2$, $-3$,
and $-4$ channels, any results based on these potentials are pure
predictions and hence will be left for a future publication. Here we
only focus on the $S=-1$ channel, for which scattering data do exist.
However, because our models heavily rely on the assumption of SU(3)
symmetry (although we allow for a breaking of this symmetry to allow
for the fact that the strange quark is much heavier than the
up and down quarks), we will here define the interaction Lagrangian,
and hence the coupling constants, for the complete baryon octet.

The eight $J^P={\textstyle\frac{1}{2}}^+$ baryons can be collected
into a traceless matrix $B$, which has the familiar form
\begin{equation}
  B = \left( \begin{array}{ccc}
      {\displaystyle\frac{\Sigma^{0}}{\sqrt{2}}+\frac{\Lambda}{\sqrt{6}}}
               &  \Sigma^{+}  &  p  \\[2mm]
      \Sigma^{-} & {\displaystyle-\frac{\Sigma^{0}}{\sqrt{2}}
                   +\frac{\Lambda}{\sqrt{6}}}  &  n \\[2mm]
      -\Xi^{-} & \Xi^{0} &  {\displaystyle-\frac{2\Lambda}{\sqrt{6}}}
             \end{array} \right),
\end{equation}
and which is invariant under SU(3) transformations. Similarly, the
various meson nonets (we take the pseudoscalar mesons with $J^P=0^+$
as an example) can be written as
\begin{equation}
     P=P_{\rm sin}+P_{\rm oct},
\end{equation}
where the singlet matrix $P_{\rm sin}$ has elements $\eta_0/\sqrt{3}$
on the diagonal, and the octet matrix $P_{\rm oct}$ is given by
\begin{equation}
   P_{\rm oct} = \left( \begin{array}{ccc}
      {\displaystyle\frac{\pi^{0}}{\sqrt{2}}+\frac{\eta_{8}}{\sqrt{6}}}
             & \pi^{+}  &  K^{+}  \\[2mm]
      \pi^{-} & {\displaystyle-\frac{\pi^{0}}{\sqrt{2}}
         +\frac{\eta_{8}}{\sqrt{6}}}  &   K^{0} \\[2mm]
      K^{-}  &  \overline{{K}^{0}}
             &  {\displaystyle-\frac{2\eta_{8}}{\sqrt{6}}}
             \end{array} \right).
\end{equation}
One can now define the SU(3)-invariant combinations
\begin{eqnarray}
  \left[\overline{B}BP\right]_{F} &=& {\rm Tr}(\overline{B}PB)
           -{\rm Tr}(\overline{B}BP)                         \nonumber\\
&=& {\rm Tr}(\overline{B}P_{\rm oct}B)-{\rm Tr}(\overline{B}BP_{\rm oct}),\\
  \left[\overline{B}BP\right]_{D} &=& {\rm Tr}(\overline{B}PB)
           +{\rm Tr}(\overline{B}BP)-{\textstyle\frac{2}{3}}\,
            {\rm Tr}(\overline{B}B){\rm Tr}(P)               \nonumber\\
&=& {\rm Tr}(\overline{B}P_{\rm oct}B)+{\rm Tr}(\overline{B}BP_{\rm oct}),\\
  \left[\overline{B}BP\right]_{S} &=&
            {\rm Tr}(\overline{B}B){\rm Tr}(P)               \nonumber\\
  &=& {\rm Tr}(\overline{B}B){\rm Tr}(P_{\rm sin}),
\end{eqnarray}
and hence an interaction Lagrangian~\cite{Swa63}
\begin{equation}
   {\cal L}_{I} = -g^{\rm oct}\sqrt{2}\left\{
     \alpha\left[\overline{B}BP\right]_{F}+
     (1-\alpha)\left[\overline{B}BP\right]_{D}\right\}\, - \,
     g^{\rm sin}{\textstyle\sqrt{\frac{1}{3}}}
     \left[\overline{B}BP\right]_{S},             \label{LIsu3}
\end{equation}
where $\alpha$ is known as the $F/(F+D)$ ratio, and the square-root
factors are introduced for later convenience.
We next introduce the isospin doublets
\begin{equation}
  N=\left(\begin{array}{c} p \\ n \end{array} \right), \ \ \
  \Xi=\left(\begin{array}{c} \Xi^{0} \\ \Xi^{-} \end{array} \right), \ \ \
  K=\left(\begin{array}{c} K^{+} \\ K^{0} \end{array} \right),
  \ \ \   K_{c}=\left(\begin{array}{c} \overline{K^{0}} \\
               -K^{-} \end{array} \right),        \label{doublets}
\end{equation}
and choose the phases of the isovector meson fields $\bbox{\Sigma}$
and $\bbox{\pi}$ such~\cite{Swa63} that
\begin{equation}
  \bbox{\Sigma}\!\cdot\!\bbox{\pi} = \Sigma^{+}\pi^{-}
       +\Sigma^{0}\pi^{0}+\Sigma^{-}\pi^{+}.
\end{equation}
If we now drop for a moment the Lorentz character of the interaction
vertices ($\gamma_{5}\gamma_{\mu}\partial^{\mu}$ for pseudoscalar
mesons), the pseudovector-coupled (derivative) pseudoscalar-meson
interaction Lagrangian is of the form
\begin{equation}
   {\cal L}_{\rm pv}={\cal L}_{\rm pv}^{\rm sin}
                    +{\cal L}_{\rm pv}^{\rm oct},
\end{equation}
where the $S$-type coupling in Eq.~(\ref{LIsu3}) gives the singlet
interaction Lagrangian
\begin{equation}
   m_{\pi}{\cal L}_{\rm pv}^{\rm sin} =
       -f_{N\!N\eta_{0}}(\overline{N}N)\eta_{0}
       -f_{\Lambda\Lambda\eta_{0}}(\overline{\Lambda}\Lambda)\eta_{0}
       -f_{\Sigma\Sigma\eta_{0}}(\overline{\bbox{\Sigma}}\!\cdot\!
        \bbox{\Sigma})\eta_{0}
       -f_{\Xi\Xi\eta_{0}}(\overline{\Xi}\Xi)\eta_{0},
                                   \label{Lbar1}
\end{equation}
with the (derivative) pseudovector coupling constants
\begin{equation}
   f_{N\!N\eta_{0}}=f_{\Lambda\Lambda \eta_{0}}
    =f_{\Sigma\Sigma\eta_{0}}=f_{\Xi\Xi \eta_{0}}
    =f^{\rm sin}_{\rm pv}.  \label{gsin}
\end{equation}
As is customary~\cite{Dum83}, we introduced the charged-pion mass as a
scaling mass to make the pseudovector coupling constants $f$ dimensionless.
The interaction Lagrangian for the meson octet is obtained by evaluating
the $F$- and $D$-type couplings in Eq.~(\ref{LIsu3}), and can be written
as
\begin{eqnarray}
   m_{\pi}{\cal L}_{\rm pv}^{\rm oct} &=&
  -f_{N\!N\pi}(\overline{N}\bbox{\tau}N)\!\cdot\!\bbox{\pi}
  +if_{\Sigma\Sigma\pi}(\overline{\bbox{\Sigma}}\!\times\!\bbox{\Sigma})
      \!\cdot\!\bbox{\pi}
  -f_{\Lambda\Sigma\pi}(\overline{\Lambda}\bbox{\Sigma}+
      \overline{\bbox{\Sigma}}\Lambda)\!\cdot\!\bbox{\pi}
  -f_{\Xi\Xi\pi}(\overline{\Xi}\bbox{\tau}\Xi)\!\cdot\!\bbox{\pi}
            \nonumber\\
 &&-f_{\Lambda N\!K}\left[(\overline{N}K)\Lambda
         +\overline{\Lambda}(\overline{K}N)\right]
   -f_{\Xi\Lambda K}\left[(\overline{\Xi}K_{c})\Lambda
         +\overline{\Lambda}(\overline{K_{c}}\Xi)\right] \nonumber\\
 &&-f_{\Sigma N\!K}\left[\overline{\bbox{\Sigma}}\!\cdot\!
         (\overline{K}\bbox{\tau}N)+(\overline{N}\bbox{\tau}K)
         \!\cdot\!\bbox{\Sigma}\right]
   -f_{\Xi\Sigma K}\left[\overline{\bbox{\Sigma}}\!\cdot\!
       (\overline{K_{c}}\bbox{\tau}\Xi)
     +(\overline{\Xi}\bbox{\tau}K_{c})\!\cdot\!\bbox{\Sigma}\right]
                                         \nonumber\\
 &&-f_{N\!N\eta_{8}}(\overline{N}N)\eta_{8}
   -f_{\Lambda\Lambda\eta_{8}}(\overline{\Lambda}\Lambda)\eta_{8}
   -f_{\Sigma\Sigma\eta_{8}}(\overline{\bbox{\Sigma}}\!\cdot\!
       \bbox{\Sigma})\eta_{8}
   -f_{\Xi\Xi\eta_{8}}(\overline{\Xi}\Xi)\eta_{8}.    \label{Lbar8}
\end{eqnarray}
The octet coupling constants are given by the following expressions
($f\equiv f^{\rm oct}_{\rm pv}$)
\begin{equation}
  \begin{array}{lll}
   f_{N\!N\pi}               = f,                               \ \ \  &
   f_{\Lambda N\!K}          =-\frac{1}{\sqrt{3}}\,f(1+2\alpha),\ \ \  &
   f_{N\!N\eta_{8}}          = \frac{1}{\sqrt{3}}\,f(4\alpha-1), \\[2mm]
      \nonumber\\
   f_{\Sigma\Sigma\pi}       = 2f\alpha,                        \ \ \  &
   f_{\Xi\Lambda K}          = \frac{1}{\sqrt{3}}\,f(4\alpha-1),\ \ \  &
   f_{\Lambda\Lambda\eta_{8}}=-\frac{2}{\sqrt{3}}\,f(1-\alpha),  \\[2mm]
      \nonumber\\
   f_{\Lambda\Sigma\pi}      = \frac{2}{\sqrt{3}}\,f(1-\alpha), \ \ \  &
   f_{\Sigma N\!K}           = f(1-2\alpha),                    \ \ \  &
   f_{\Sigma\Sigma\eta_{8}}  = \frac{2}{\sqrt{3}}\,f(1-\alpha),  \\[2mm]
      \nonumber\\
   f_{\Xi\Xi\pi}             =-f(1-2\alpha),                    \ \ \  &
   f_{\Xi\Sigma K}           =-f,                               \ \ \  &
   f_{\Xi\Xi\eta_{8}}        =-\frac{1}{\sqrt{3}}\,f(1+2\alpha).
  \end{array}
                        \label{goct}
\end{equation}
Similar relations (without the scaling mass $m_{\pi}$) are found
for the coupling constants of the scalar and vector mesons.

The assumption of SU(3) symmetry thus implies that for each type of
meson (pseudoscalar, vector, scalar) we need only four parameters to
characterize their couplings with all possible baryons: the singlet
coupling constant, the octet coupling constant, the $F/(F+D)$ ratio,
and a mixing angle which relates the physical isoscalar mesons to
their pure octet and singlet counterparts.
However, it is not {\it a priori\/} obvious that these SU(3) relations
for the coupling constants will be satisfied exactly. For example, the
strange quark is much heavier than the up and down quarks, and so
already on the quark-mass level the SU(3) symmetry is clearly broken.

In our models, breaking of the SU(3) symmetry is introduced in several
places as well. First of all, we use the physical masses for the baryons
and mesons. Second, we allow for the fact that the $\Lambda$ and
$\Sigma^0$ have the same quark content, and so there is an appreciable
mixing between the isospin-pure $\Lambda$ and $\Sigma^0$
states~\cite{Dal64}. Although exact SU(3) symmetry requires that
$f_{\Lambda\Lambda\pi^0}=0$, $\Lambda$-$\Sigma^0$ mixing and the
interaction $\Sigma^0\rightarrow\Lambda+\pi^0$ result in a non-zero
pion coupling constant for the physical $\Lambda$-hyperon. Dalitz and
von Hippel derive~\cite{Dal64}
\begin{equation}
  f_{\Lambda\Lambda\pi}=-2\frac{\langle\Sigma^0|\delta M|\Lambda\rangle}
             {M_{\Sigma^0}-M_{\Lambda}}\,f_{\Lambda\Sigma\pi},
\end{equation}
where the $\Sigma\Lambda$ element of the electromagnetic mass matrix
is given by
\begin{equation}
   \langle\Sigma^0|\delta M|\Lambda\rangle=
       \left[M_{\Sigma^0}-M_{\Lambda}+M_p-M_n\right]/\sqrt{3}.
\end{equation}
Substituting for the physical baryon masses, we find
\begin{equation}
  f_{\Lambda\Lambda\pi}=-0.0283\,f_{\Lambda\Sigma\pi}.
\end{equation}
Writing out the nucleon-nucleon-pion part of the interaction Lagrangian
(\ref{Lbar8}), we find
\begin{equation}
    (\overline{N}\bbox{\tau}N)\!\cdot\!\bbox{\pi} =
    \overline{p}p\pi^0-\overline{n}n\pi^0+\sqrt{2}\,\overline{p}n\pi^+
                      +\sqrt{2}\,\overline{n}p\pi^-,
\end{equation}
and so the neutral pion is seen to couple to the neutron with opposite
sign as compared to its coupling to the proton. This implies that the
non-zero $f_{\Lambda\Lambda\pi^0}$ coupling produces strong deviations
from charge symmetry for the $\Lambda p$ and $\Lambda n$ potentials.
Obviously, $\Lambda$-$\Sigma^0$ mixing also gives non-zero
$\Lambda\Lambda$ coupling constants for the other neutral isovector
mesons, but they give rise to much smaller effects.

Finally, we use the $^3P_0$ model~\cite{Mic69,Rijxx} to account for
the fact that the strange quark is much heavier than the up and down
quarks. In this model, the breaking of the SU(3)-flavor symmetry is
described by one parameter $\lambda_{\rm fsb}$, where we allow for a
different parameter for each meson nonet. This will be discussed in
more detail in Sec.~\ref{sec:su3break}.

In order to define the potential in momentum space, we next consider
the general baryon-baryon scattering reaction
\begin{equation}
   B_1(p_1)+B_2(p_2) \rightarrow B_3(p_3)+B_4(p_4),
\end{equation}
where the four-momentum of baryon $B_i$ is $p_i=(E_i,{\bf p}_i)$,
with $E_i=\sqrt{{\bf p}^2_i+M^2_i}$ and $M_i$ its mass.
The second-order one-meson-exchange kernel is derived following the
procedure as discussed in our earlier papers on two-meson
exchange~\cite{Rij92,Rij96}, to which we refer for details and
definitions. In this procedure the Thompson equation~\cite{Tho70}
for the wave function reads
\begin{equation}
   \phi_{++}({\bf p}')=\phi_{++}^{(0)}({\bf p}')
         +E_2^{(+)}({\bf p}';W)
          \int d^3pK^{\rm irr}({\bf p}',{\bf p}|W)
          \phi_{++}({\bf p}),          \label{Thoeq}
\end{equation}
with $W=\sqrt{s}$ the total energy, and ${\bf p}$ and ${\bf p}'$ the
center-of-mass momenta in the initial and final states, respectively.
The irreducible kernel is given by
\begin{eqnarray}
   K^{\rm irr}({\bf p}',{\bf p}|W) &=& -(2\pi)^2
    \left[W-E_3({\bf p}')-E_4({\bf p}')\right]\,
    \left[W-E_1({\bf p})-E_2({\bf p})\right]\,
    \int_{-\infty}^{\infty}\!\!dp'_0
    \int_{-\infty}^{\infty}\!\!dp_0              \nonumber\\
   &\times & \left\{\left[F_W^{(3)}({\bf p}',p'_0)
                          F_W^{(4)}(-{\bf p}',-p'_0)\right]^{-1}
                    \left[I({\bf p}',p'_0;{\bf p},p_0)\right]_{++,++}
                    \left[F_W^{(1)}({\bf p},p_0)
                          F_W^{(2)}(-{\bf p},-p_0)\right]^{-1}\right\}.
\end{eqnarray}
Substituting for the one-meson-exchange Feynman propagator and
performing the $p_0$ and $p'_0$ integrations generates the two
time-ordered one-meson-exchange diagrams with energy denominator
\begin{equation}
   D(\omega)=\frac{1}{2\omega}\left[\frac{1}{E_2+E_3-W+\omega}
             +\frac{1}{E_1+E_4-W+\omega}\right].
\end{equation}
Here, $\omega^2={\bf k}^2+m^2$, with $m$ the meson mass and
${\bf k}={\bf p}'-{\bf p}$ the momentum transfer.
In the static approximation $E_i\rightarrow M_i$ and
$W\rightarrow M^0_1+M^0_2$. Note that we have included a superscript 0
to indicate that these masses refer to the masses of the particular
interaction channel we are considering. They are not necessarily
equal to the masses $M_1$ and $M_2$ occurring in the time-ordered
diagrams.
For example, the potential for the $\Sigma N\rightarrow\Sigma N$
contribution in the coupled-channel $(\Lambda N,\Sigma N)$ system
has $M_1=M_{\Sigma}$ and $M_2=M_N$, but $M^0_1=M_{\Lambda}$ and
$M^0_2=M_N$.

In principle, the propagator in the static approximation can be
handled exactly using the fact that~\cite{Rij92}
\begin{equation}
  \frac{1}{\omega(\omega+a)}=\frac{2}{\pi}\int_0^{\infty}
          \frac{ad\lambda}{(\omega^{2}+\lambda^2)(a^2+\lambda^2)}
          +\frac{2\theta(-a)}{\omega^2-a^2},\ \ \ (a<m). \label{oma}
\end{equation}
However, this requires an additional (numerical) evaluation of an
integral whenever $a\neq0$, which might be a considerable time
factor in practical calculations. A way to avoid this additional
integral is to assume that the average of the initial and final masses
always approximately equals the mass of the interaction channel,
$M^0_1+M^0_2$. The advantage of this, more crude, approximation is
that the propagator can then be written as
\begin{equation}
   D(\omega)\rightarrow \frac{1}{\omega^2-{\textstyle\frac{1}{4}}
            (M_3-M_4+M_2-M_1)^2},       \label{Deffmass}
\end{equation}
which means we have introduced an effective meson mass $\overline{m}$,
where the mass has dropped to
\begin{equation}
    m^2 \rightarrow \overline{m}^2=m^2-{\textstyle\frac{1}{4}}
                 (M_3-M_4+M_2-M_1)^2.
\end{equation}
The change in mass can be considerable for certain potentials. For
example, the effective kaon mass in $\Sigma N\rightarrow N\Sigma$
drops from 495.8 MeV/$c^2$ to 425.8 MeV/$c^2$. In the following, we
will use the static approximation in the form of Eq.~(\ref{Deffmass}).
In view of the relatively large error bars on the experimental $Y\!N$
scattering data, we argue that at present it is not worthwhile to
pursue the more complicated exact treatment; we leave this for a
later study. Note also that this approximation still ensures that
the potential, viewed as a matrix in channel space, is symmetric,
as required by time-reversal invariance.

The transition from the Thompson equation (\ref{Thoeq}) to the
Lippmann-Schwinger equation,
\begin{equation}
   \phi({\bf p}')=\phi^{(0)}({\bf p}')+g({\bf p}';W)
        \int\!\!d^3pV({\bf p}',{\bf p}|W)\phi({\bf p}),  \label{LSeq}
\end{equation}
is made by defining the transformations
\[
   \phi_{++}({\bf p})=N({\bf p};W)\phi({\bf p}),
\]
\[
   K^{\rm irr}({\bf p}',{\bf p}|W)=N^{-1}({\bf p}';W)
        V({\bf p}',{\bf p}|W)N^{-1}({\bf p};W),
\]
\begin{equation}
     E_2^{(+)}({\bf p};W)=N^2({\bf p};W)g({\bf p};W),
\end{equation}
with the Green's function
\begin{equation}
   g({\bf p};W)=\frac{1}{(2\pi)^3}\Lambda_+^{(1)}({\bf p})
             \Lambda_+^{(2)}({\bf p})
             \frac{2M_{\rm red}}{{\bf p}_i^2-{\bf p}^2+i\delta},
\end{equation}
with $\Lambda_+({\bf p})$ a spin-projection operator and ${\bf p}_i$
the on-shell momentum associated with $W$. This defines the potential.
We make the standard expansions and approximations valid for
low-energy scattering and end up with the potentials as given in
Ref.~\cite{Mae89}\footnote{Note that in Ref.~\protect\cite{Mae89},
$\Omega_2^{(P)}$ in Eq.~(23) should have a minus sign and
$\Omega_6^{(S)}$ in Eq.~(25) should have masses squared in the
denominator.}.
The partial-wave projection for the momentum-space potential is
discussed in Ref.~\cite{Rij95}.

The potentials are regularized with a Gaussian cut off, which still
allows for the Fourier transform to configuration space to be carried
out analytically. Details again can be found in Ref.~\cite{Mae89}.
Unfortunately, this reference contains a number of typographical
errors. The corrected expressions are given in the appendix, where
the potentials refer to the scattering process where one of the meson
vertices occurs between $B_1$ and $B_3$, and the other between $B_2$
and $B_4$. The mass $M_{13}$ then denotes the average of the $B_1$ and
$B_3$ masses, and $M_{24}$ the average of the $B_2$ and $B_4$ masses.
For the exchanged diagram we have to interchange $3\leftrightarrow4$
everywhere and multiply by the exchange operator $P$.
The exchange operator $P=+1$ for even-$L$ singlet and odd-$L$
triplet partial waves, and $P=-1$ for odd-$L$ singlet and even-$L$
triplet partial waves. For $Y\!N$ scattering, the exchanged diagram
only occurs when the exchanged meson carries strangeness ($K$, $K^*$,
$\kappa$, $K^{**}$).

\section{$\protect\bbox{Y\!N}$ channels}
\label{sec:YNchan}
In our approach, the potentials are calculated on the isospin basis.
Because the two nucleons form an isodoublet, the $\Lambda$-hyperon
an isosinglet, and the three $\Sigma$-hyperons an isotriplet, there
are only two isospin channels:
\begin{eqnarray}
   I={\textstyle\frac{1}{2}}:\ \ && (\Lambda N,\Sigma N)\rightarrow
                                  (\Lambda N,\Sigma N),     \nonumber\\
   I={\textstyle\frac{3}{2}}:\ \ && \Sigma N\rightarrow\Sigma N.
\end{eqnarray}
The isospin factors for the various meson exchanges in the two
isospin channels are given in Table~\ref{tabisofac}. We use the
pseudoscalar mesons as a specific example, and $P$ is the exchange
operator alluded to in the previous section. We also include the
coupling of the $\Lambda$-hyperon to the neutral pion, which is
non-zero due to $\Lambda$-$\Sigma^0$ mixing, as was discussed earlier.
However, this matrix element is {\it only\/} included when the
potentials are used for calculations on the physical particle basis.

In the physical particle basis, there are four charge channels:
\begin{eqnarray}
   q=+2:\ \  && \Sigma^+p\rightarrow\Sigma^+p,         \nonumber\\
   q=+1:\ \  && (\Lambda p, \Sigma^+n, \Sigma^0p)\rightarrow
             (\Lambda p, \Sigma^+n, \Sigma^0p),        \nonumber\\
   q=\,\ \ 0:\ \ && (\Lambda n, \Sigma^0n, \Sigma^-p)\rightarrow
             (\Lambda n, \Sigma^0n, \Sigma^-p),        \nonumber\\
   q=-1:\ \  && \Sigma^-n\rightarrow\Sigma^-n.
\end{eqnarray}
Obviously, the potential on the particle basis for the $q=2$ and
$q=-1$ channels are given by the $I={\textstyle\frac{3}{2}}$
$\Sigma N$ potential on the isospin basis, substituting the appropriate
physical particle masses. For $q=1$ and $q=0$, the potentials are related
to the potentials on the isospin basis by an isospin rotation.
Using a notation where we only list the hyperons
[$V_{\Lambda\Sigma^+}=(\Lambda p|V|\Sigma^+n)$, etc.], we find for $q=1$
\begin{equation}
  \left(\begin{array}{ccc}
    V_{\Lambda\Lambda}  & V_{\Lambda\Sigma^+}  & V_{\Lambda\Sigma^0} \\[2mm]
    V_{\Sigma^+\Lambda} & V_{\Sigma^+\Sigma^+} & V_{\Sigma^+\Sigma^0}\\[2mm]
    V_{\Sigma^0\Lambda} & V_{\Sigma^0\Sigma^+} & V_{\Sigma^0\Sigma^0}
        \end{array}\right) =
  \left(\begin{array}{ccc}
    V_{\Lambda\Lambda}
    &  \sqrt{\textstyle\frac{2}{3}}V_{\Lambda\Sigma}
    & -\sqrt{\textstyle\frac{1}{3}}V_{\Lambda\Sigma} \\[2mm]
    \sqrt{\textstyle\frac{2}{3}}V_{\Sigma\Lambda}
    &  {\textstyle\frac{2}{3}}V_{\Sigma\Sigma}({\textstyle\frac{1}{2}})
      +{\textstyle\frac{1}{3}}V_{\Sigma\Sigma}({\textstyle\frac{3}{2}})
    &  {\textstyle\frac{1}{3}}\sqrt{2}\left[
           V_{\Sigma\Sigma}({\textstyle\frac{3}{2}})
          -V_{\Sigma\Sigma}({\textstyle\frac{1}{2}})\right] \\[2mm]
    -\sqrt{\textstyle\frac{1}{3}}V_{\Sigma\Lambda}
    &  {\textstyle\frac{1}{3}}\sqrt{2}\left[
           V_{\Sigma\Sigma}({\textstyle\frac{3}{2}})
          -V_{\Sigma\Sigma}({\textstyle\frac{1}{2}})\right]
    &  {\textstyle\frac{1}{3}}V_{\Sigma\Sigma}({\textstyle\frac{1}{2}})
      +{\textstyle\frac{2}{3}}V_{\Sigma\Sigma}({\textstyle\frac{3}{2}})
        \end{array}\right),
\end{equation}
while for $q=0$ we find
\begin{equation}
  \left(\begin{array}{ccc}
    V_{\Lambda\Lambda}  & V_{\Lambda\Sigma^0}  & V_{\Lambda\Sigma^-} \\[2mm]
    V_{\Sigma^0\Lambda} & V_{\Sigma^0\Sigma^0} & V_{\Sigma^0\Sigma^-}\\[2mm]
    V_{\Sigma^-\Lambda} & V_{\Sigma^-\Sigma^0} & V_{\Sigma^-\Sigma^-}
        \end{array}\right) =
  \left(\begin{array}{ccc}
    V_{\Lambda\Lambda}
    &  \sqrt{\textstyle\frac{1}{3}}V_{\Lambda\Sigma}
    & -\sqrt{\textstyle\frac{2}{3}}V_{\Lambda\Sigma} \\[2mm]
    \sqrt{\textstyle\frac{1}{3}}V_{\Sigma\Lambda}
    &  {\textstyle\frac{1}{3}}V_{\Sigma\Sigma}({\textstyle\frac{1}{2}})
      +{\textstyle\frac{2}{3}}V_{\Sigma\Sigma}({\textstyle\frac{3}{2}})
    &  {\textstyle\frac{1}{3}}\sqrt{2}\left[
           V_{\Sigma\Sigma}({\textstyle\frac{3}{2}})
          -V_{\Sigma\Sigma}({\textstyle\frac{1}{2}})\right] \\[2mm]
    -\sqrt{\textstyle\frac{2}{3}}V_{\Sigma\Lambda}
    &  {\textstyle\frac{1}{3}}\sqrt{2}\left[
           V_{\Sigma\Sigma}({\textstyle\frac{3}{2}})
          -V_{\Sigma\Sigma}({\textstyle\frac{1}{2}})\right]
    &  {\textstyle\frac{2}{3}}V_{\Sigma\Sigma}({\textstyle\frac{1}{2}})
      +{\textstyle\frac{1}{3}}V_{\Sigma\Sigma}({\textstyle\frac{3}{2}})
        \end{array}\right).
\end{equation}

The relativistic relation between the on-shell center-of-mass momentum
$p_i$ in channel $i$ and the total energy $\sqrt{s}$ is given by
\begin{equation}
   p^2_i=\frac{1}{4s}\left[s-(M_1(i)+M_2(i))^2\right]
                     \left[s-(M_1(i)-M_2(i))^2\right], \label{pcmrel}
\end{equation}
while the total energy squared for a specific interaction channel $i$
with laboratory momentum $p_{\rm lab}(i)$ is given by
\begin{equation}
   s=M_1^2(i)+M_2^2(i)+2M_2(i)\sqrt{p_{\rm lab}^2(i)+M_1^2(i)}.
                                                      \label{plabrel}
\end{equation}
Expanding the square-root energies, we obtain the corresponding
nonrelativistic expressions:
\begin{eqnarray*}
   && p_i^2=2M_{\rm red}(i)\left[\sqrt{s}-M_1(i)-M_2(i)\right], \\
   && \sqrt{s}=M_1(i)+M_2(i)+M_{\rm red}(i)
               \left[p_{\rm lab}^2(i)/2M_1^2(i)\right].
\end{eqnarray*}
We always use the relativistic relations (\ref{pcmrel}) and
(\ref{plabrel}). Substituting for the empirical baryon masses, the
various $\Sigma N$ thresholds in the $\Lambda p$ channel are found
to be at
\begin{equation}
   p_{\rm lab}^{\rm th}(\Lambda p\rightarrow\Sigma^+n)
                  =633.4 \mbox{ MeV}/c,\ \ \ \
   p_{\rm lab}^{\rm th}(\Lambda p\rightarrow\Sigma^0p)
                  =642.0 \mbox{ MeV}/c;
\end{equation}
those in the $\Lambda n$ channel at
\begin{equation}
   p_{\rm lab}^{\rm th}(\Lambda n\rightarrow\Sigma^0n)
                  =641.7 \mbox{ MeV}/c,\ \ \ \
   p_{\rm lab}^{\rm th}(\Lambda n\rightarrow\Sigma^-p)
                  =657.9 \mbox{ MeV}/c;
\end{equation}
and the average (single) threshold for the potential on the
isospin basis at
\begin{equation}
    p_{\rm lab}^{\rm th}(\Lambda N\rightarrow\Sigma N)
    =643.8 \mbox{ MeV}/c.
\end{equation}
Using nonrelativistic kinematics, the thresholds are found to be
lower by about 30 MeV/$c$.

There are various ways to solve the Lippmann-Schwinger equation for
the partial-wave momentum-space potential. We use the Kowalski-Noyes
method~\cite{Noy65,Kow65} to handle the singularities in the Green's
function for the open channels. The Coulomb interaction in the
$\Sigma^+p\rightarrow\Sigma^+p$ and $\Sigma^-p\rightarrow\Sigma^-p$
channels is included via the Vincent-Phatak method~\cite{Vin74}.

The multichannel Schr\"odinger equation for the configuration-space
potential is derived from the Lippmann-Schwinger equation through
the standard Fourier transform, and the equation for the partial-wave
radial wave function is found to be of the form~\cite{Mae89}
\begin{equation}
   u^{\prime\prime}_{l,j}+(p_i^2\delta_{ij}-A_{ij})u_{l,j}
       -B_{ij}u'_{l,j}=0,           \label{radialeq}
\end{equation}
where $A_{ij}$ contains the potential, nonlocal contributions, and
the centrifugal barrier, while $B_{ij}$ is only present when nonlocal
contributions are included. This equation can be easily solved
numerically using a method derived by Bergervoet~\cite{Ber88}.
A discussion of how to handle the presence of closed channels is given,
for example, in Ref.~\cite{Nag73}.
As is well known, the inclusion of the Coulomb interaction in the
configuration-space equation poses no additional complications.

The potentials are of such a form that they are exactly equivalent
in both momentum space and configuration space. This means that the
resulting phase shifts and mixing parameters are also the same,
provided both equations (\ref{LSeq}) (in the static approximation)
and (\ref{radialeq}) are solved with sufficient accuracy.

\section{Mesons, coupling constants, and flavor SU(3)}
\label{sec:mesons}
\subsection{The pseudoscalar mesons $\protect\bbox{J^{PC}=0^{-+}}$}
In the literature one encounters two couplings for the pseudoscalar mesons
to the $J^{P}={\textstyle\frac{1}{2}}^{+}$ baryons: the pseudoscalar
coupling, ${\cal L}_{ps}=g\overline{\psi}i\gamma_{5}\psi\phi$, and the
pseudovector coupling, ${\cal L}_{pv}=(f/m_{\pi})\overline{\psi}\gamma_5
\gamma_{\mu}\psi\partial^{\mu}\phi$ (or a mixture of these two).
We assume SU(3) for the pseudovector coupling $f$. Then, the Cabibbo
theory of the weak interactions and the Goldberger-Treiman relation give
$\alpha_{PV}=[F/(F+D)]_{pv}=0.355$~\cite{Dum83}.
In the Nijmegen soft-core models, this value could be imposed while
still keeping an excellent description of the $Y\!N$ data, including the
accurate datum on the capture ratio at rest.

The Nijmegen soft-core OBE models have quite sizable couplings to the
baryons for the scalar $\varepsilon$ meson (see below). If this were to
be used in a model for the pion-nucleon interaction together with the
pseudovector coupling for the pion, one would expect a large violation
of the soft-pion constraints on the $\pi N$ scattering lengths. However,
the Nijmegen soft-core OBE models are compatible with these soft-pion
constraints, because the potentially dangerous $\varepsilon$ contribution
is canceled by an opposite pomeron-exchange contribution~\cite{Swa89}.

\subsection{The vector mesons $\protect\bbox{J^{PC}=1^{--}}$}
An important ingredient of the baryon-baryon interaction is the exchange
of the members of the vector-meson nonet $(\rho$, $\phi$, $\omega$,
$K^{*})$. The details of our treatment of the vector mesons have been
given in Refs.~\cite{Nag75,Nag77,Nag79}; see also \cite{Swa93}.
Ideal mixing between $\omega$ and $\phi$ implies $\theta_V=35.3^\circ$,
which means that the $\phi$ meson would be pure $s\bar{s}$, and hence
would not couple to the nucleon. We assume a small deviation from ideal
mixing and use the experimental value $\theta_V=37.5^\circ$~\cite{Dum83}.
For the electric $F/(F+D)$ ratio we take $\alpha_V^e=1$, as required
by the ``universality'' assumption~\cite{Sak60}.
The magnetic $\alpha_V^m$ is not always the same. In the OBE models,
the singlet-triplet strength in $\Lambda N$ depends, besides on other
things, especially on $\alpha_V^m$.
This feature is used to construct a range of soft-core models.

\subsection{The scalar mesons $\protect\bbox{J^{PC}=0^{++}}$}
The scalar mesons have constituted an important role in the construction
of the Nijmegen potential models since 1970. They are an essential
ingredient both in the hard-core models D~\cite{Nag77} and
F~\cite{Nag79}, and in all the soft-core models as well.

The scalar meson $\sigma(550)$ was introduced in 1960-1962 by Hoshizaki
{\it et al.}~\cite{Hos60}. In the OBE models for $N\!N$, this scalar
meson was necessary for providing sufficient intermediate-range central
attraction and for the spin-orbit interaction required to describe
the $^{3}P_{J}$ splittings. In 1971 it was realized that the exchange
of the broad $\varepsilon(760)$ could explain the role of the fictitious
$\sigma$ meson~\cite{Sch71,Bin71}. {}From then on, this broad
$\varepsilon(760)$ has been used in the Nijmegen OBE models.
A recent analysis of $\pi$ production in $\pi N$ scattering with polarized
nucleons claimed to have found unambiguous evidence for a broad isoscalar
$J^{PC}=0^{++}$ state under the peak of the $\rho$ meson~\cite{Sve92a}.
This was based on an amplitude analysis involving besides $\pi$ exchange
also $a_{1}$ exchange in the production mechanism. In a similar analysis
of data on $K^{+}n\rightarrow K^{+}\pi^{-}p$, evidence was found for an
$I={\textstyle\frac{1}{2}}$, $0^{+}(887)$ strange scalar meson under the
peak of the $K^{*}(892)$ meson~\cite{Sve92b}. In the latest issue of
the Particle Data Group~\cite{PDG96} this analysis is cited with reserve,
asserting that the $\varepsilon$ parameters of \cite{Sve92a} cannot be
correct because the $f_{0}(980)$ is neglected in the analysis.

Gilman and Harrari~\cite{Gil68} showed that all Adler-Weisberger sum
rules can be satisfied by saturation in the mesonic sector with the
$\pi(140)$, $\varepsilon(760)$, $\rho(760)$, and $a_{1}(1090)$.
They found the $\varepsilon$, in \cite{Gil68} called $\sigma$, to be
degenerate with the $\rho$, having a width of $\Gamma(\varepsilon
\rightarrow\pi\pi)=570$ MeV. Used in this work were the Regge high-energy
behavior, SU(2)$\otimes$SU(2) chiral algebra of charges, and pion
dominance of the divergence of the axial-vector current. Similar
phenomenology was derived by Weinberg requiring that the sum of the
tree graphs for forward pion scattering, generated by a chiral-invariant
Lagrangian, should not grow faster at high energies than permitted by
Regge behavior of the actual amplitudes~\cite{Wei68,Wei90}. Therefore,
it seems that chiral symmetry combined with Regge behavior requires
a broad scalar $\varepsilon$ degenerate with the $\rho$.
Finally, we should mention that the Helsinki group now also finds an
$\varepsilon$ meson and other members of a scalar nonet~\cite{Tor95}.

In the quark model, the scalar mesons have been viewed as conventional
$^{3}P_{0}$ $q\bar{q}$ states, while others view them as crypto-exotic
$q^{2}\bar{q}^{2}$ states~\cite{Jaf77} or glueball states. We will
briefly review the assignments as $q\bar{q}$ and as $q^{2}\bar{q}^{2}$
states.

In the $q\bar{q}$ picture, one has for the unitary singlet and octet
states, denoted respectively by $\varepsilon_{1}$ and $\varepsilon_{8}$,
\begin{eqnarray}
  \varepsilon_{1} &=& \left(u\bar{u}+d\bar{d}+s\bar{s}\right)/\sqrt{3},
                      \nonumber\\
  \varepsilon_{8} &=& \left(u\bar{u}+d\bar{d}-2s\bar{s}\right)/\sqrt{6}.
\end{eqnarray}
The physical states are mixings of the pure SU(3) states and we write
\begin{eqnarray}
 \varepsilon &=& \cos\theta_S\varepsilon_{1}
           +\sin\theta_S\varepsilon_{8}, \nonumber\\
 f_{0}  &=&-\sin\theta_S\varepsilon_{1}
           +\cos\theta_S\varepsilon_{8}.
\end{eqnarray}
Then, for ideal mixing we have $\tan\theta_S=1/\sqrt{2}$ or
$\theta_S\approx 35.3^\circ$, and so
\begin{eqnarray}
   \varepsilon &=& f_{0}(760)=(u\bar{u}+d\bar{d})/\sqrt{2}, \nonumber\\
   f_{0}  &=& f_{0}(980)=s\bar{s}.
\end{eqnarray}
Note that in contrast to \cite{Mae89}, we here follow for the
description of the meson mixing the same conventions as for the
pseudoscalar and vector mesons.

In the $q^{2}\bar{q}^{2}$ picture~\cite{Jaf77} (see also \cite{Swa93}),
one introduces diquarks $q^{2}$ with $F=3^{*}$, $C=3^{*}$, and $S=0$,
for the flavor, color, and spin representations, respectively.
Since $F=3^{*}$, one denotes these diquark states by $\overline{Q}$.
This conjugated triplet $\overline{Q}$ has the contents
$\overline{S}=[ud]$, $\overline{U}=[sd]$, and $\overline{D}=[su]$,
where $[ud]$ stands for the antisymmetric flavor wave function $ud-du$,
and so on. The $Q\overline{Q}$ states form a scalar flavor nonet.
In particular, Jaffe predicted the lowest-mass state (which we assume
here to be $\varepsilon$) as $S\overline{S}$, with $I=0$,
$J^{PC}=0^{++}$, and mass $M=690$ MeV. In this scalar nonet, Jaffe
predicted a degenerate pair of $I=0$ and $I=1$ state at $M=1150$ MeV.
It seems natural to identify these with the $f_{0}(980)$ and the
$a_{0}(980)$.
Explicitly, in the $q^{2}\bar{q}^{2}$ model, the quark content of
the neutral states $f_{0}(760)$, $f_{0}(980)$, and $a_{0}(980)$ is
\begin{eqnarray}
  && S\overline{S} = [\bar{u}\bar{d}][ud], \nonumber\\
  && \left(U\overline{U} \pm D\overline{D}\right) =
     \left\{[\bar{s}\bar{d}][sd] \pm [\bar{s}\bar{u}][su]\right\}/\sqrt{2}.
\end{eqnarray}
The strange members of this nonet are combinations like $\kappa^{+}\sim
[ud][\bar{s}\bar{d}]$, etc. These are expected at about $M=880$ MeV,
just under the $K^{*}(892)$.
Ideal mixing in the case of the $q^{2}\bar{q}^{2}$ states means that
\begin{eqnarray}
   \varepsilon &=& f_{0}(760)=S\overline{S}, \nonumber\\
   f_{0}  &=& f_{0}(980)=(U\overline{U}+D\overline{D})/\sqrt{2},
\end{eqnarray}
which in this case implies that $\tan\theta_S=-\sqrt{2}$, or
$\theta_S\approx-54.8^\circ$.

In view of the above, we note that ideal mixing for the scalar mesons
in the case of $q^{2}\bar{q}^{2}$ states is quite distinct from that for
the $q\bar{q}$ states. To analyze some of the differences between the
$q\bar{q}$ and the $q^{2}\bar{q}^{2}$ assignments for the $BB$ channels,
we remind the reader that in our strategy we keep the $N\!N$ channel
fixed. Considering the mixing, one obtains for $g_{N\!N\varepsilon}$
and $g_{N\!Nf_0}$, in terms of the flavor singlet and octet couplings,
\begin{eqnarray}
   g_{N\!N\varepsilon} &=& \cos\theta_S g_{1}+\sin\theta_S g_{8},
                            \nonumber\\
   g_{N\!Nf_0} &=&-\sin\theta_S g_{1}+\cos\theta_S g_{8},
\end{eqnarray}
where $g_{1}=g_{N\!N\varepsilon_1}$ and $g_{8}=g_{N\!N\varepsilon_8}=
(4\alpha_S-1)g_{N\!Na_0}/\sqrt{3}$. Because $g_{N\!Na_0}$,
$g_{N\!N\varepsilon}$, and $g_{N\!Nf_0}$ are fitted to the $N\!N$
scattering data, the only freedom left for the $Y\!N$ and the $YY$
systems is in the variation of the scalar mixing angle $\theta_S$.
The scalar $F/(F+D)$ ratio is restricted by
\begin{equation}
   g_{8} \equiv \frac{(4\alpha_S-1)}{\sqrt{3}}g_{N\!Na_0} =
   \sin\theta_S g_{N\!N\varepsilon} + \cos\theta_S g_{N\!Nf_0},
                                 \label{alphas}
\end{equation}
from which it is clear that $\alpha_S=\alpha_S(\theta_S)$.
This relation implies roughly that for positive values of $\theta_S$
we get $\alpha_S>0$, while for negative values we get $\alpha_S<0$.
For the ideal mixing in the $q\bar{q}$ case $\alpha_S\approx+1.0$, and
for ideal mixing in the $q^{2}\bar{q}^{2}$ case $\alpha_S\approx-1.0$.
This difference between the $q\bar{q}$ and the $q^{2}\bar{q}^{2}$
assignment is quite important for the $Y\!N$ and the $YY$ systems.
In principle, one could of course allow for the possibility that the
actual physical states, $\varepsilon(760)$ and $f_{0}(980)$, are
mixtures of the $q\bar{q}$ and the $q^{2}\bar{q}^{2}$ states. We expect
that $\theta_S>0$ if the $q\bar{q}$ component dominates, whereas
$\theta_S<0$ when the $q^{2}\bar{q}^{2}$ component dominates.

In Fig.~\ref{fig:scalYN}, we show the strength of the scalar-exchange
central potential, in arbitrary units, for the diagonal matrix elements
in $Y\!N$. Here, we assumed equal masses for the members of the scalar
nonet. Considering the contribution from the scalar nonet, we note the
following. In the $\Sigma^{+}p(^{3}S_{1})$ channel, the scalar-nonet
contribution is attractive in the $q\bar{q}$ case, whereas in the
$q^{2}\bar{q}^{2}$ case it is repulsive.
Note that for the spin-singlet the interaction in $\Lambda\!N$
is quite similar to that in $\Sigma N$, due to the dominance of the
$\{27\}$ irrep.
Although outside the scope of the present paper, we mention that in
the $\Lambda\Lambda(^{3}S_{1})$ channel the scalar-nonet contribution
is much stronger for $q^{2}\bar{q}^{2}$ domination than for $q\bar{q}$
domination. A similar situation occurs for the $\Xi N(^{1}S_{0},I=0)$
and $\Xi N(^{3}S_{1},I=1)$ states.
So far, the soft-core OBE models all have $\theta_S>30^\circ$, which
indeed implies that the $\Lambda\Lambda$ and the $\Xi N$ potentials are
rather weakly attractive in the intermediate range. They therefore
cannot produce sufficient attraction to account for the binding energies
of the experimentally found double-Lambda hypernuclei, e.g.,
$_{\Lambda\Lambda}^{10}$Be~\cite{Dal89}.

\subsection{The pomeron $\protect\bbox{J^{PC}=0^{++}}$
            and the heavy mesons}
The physical nature of pomeron-exchange can be understood in the
framework of QCD as a two-gluon (or multigluon) exchange effect.
In the Low-Nussinov two-gluon model~\cite{Low75}, it was once
proposed~\cite{Pum81} to distribute the two-gluon coupling over the
quarks of a hadron, the so-called ``subtractive pomeron''.
Then, one would expect at low energies an attractive van der Waals
type of force. This is in conflict with the results from Regge
phenomenology~\cite{Rij85}. However, it became apparent experimentally
in the study of the $pp\rightarrow(\Lambda\phi K^{+})p$ and
$pp\rightarrow(\Lambda\overline{\Lambda}p)p$ reactions at $\sqrt{s}=63$
GeV~\cite{Smi85,Hen92} that the pomeron couples dominantly to
individual quarks. This leads to the so-called ``additive pomeron''.
The dominance of the one-quark coupling can be understood as due to the
fact that in the case of a coupling to two quarks the loop momentum
involved in such a coupling has to pass through at least one baryon.
Thus, the baryon wave function is involved, which leads to a suppression
of $a^{2}/R^{2}$~\cite{Lan87}, where $a$ and $R$ are the quark and baryon
radius, respectively.
It is interesting to know whether this is also true at lower energies.
In the Low-Nussinov model one can argue that the pomeron-quark coupling
leads to a repulsive Gaussian potential~\cite{Swa93}, which has been
used in the Nijmegen soft-core models. The importance of the pomeron
in OBE models being compatible with chiral symmetry has already been
mentioned above, see also \cite{Swa89}.

Exact SU(3) and unitarity cause a strong mixing between the ``bare''
pomeron and the isosinglet member of the tensor mesons. Medium strong
SU(3) breaking then gives mixing of these bare states, leading to the
physical pomeron and $f_2$ tensor meson. This is why we include the
$J=0$ contributions from the tensor $f_2$, $f'_2$, and $a_2$ Regge
trajectories.
So far, the explicit exchange of axial and tensor mesons has hardly
been explored in models of baryon-baryon interactions for low energies.
The axial mesons are very important in connection with chiral symmetry
and play an important role in sum rules~\cite{Wei67}. The tensor mesons
are very important at higher energies, lying on a dominant Regge
trajectory, and they are exchange-degenerate with the vector mesons.
In principle, there is no problem in the present approach to incorporate
these heavy mesons. (We already include the $J=0$ contribution from
the tensor mesons.)
Recently, we have included these mesons explicitly, using the estimates
based on the Regge hierarchy from \cite{Rij85} as a guidance for the
coupling constants. With regard to the general features, no qualitative
changes in the description of the $N\!N$ and $Y\!N$ channels were
observed. This can be understood from the fact that these mesons have
masses well above 1 GeV, and hence are expected to affect the interaction
only at very short distances. But the short-distance part of the
interaction can already very well be parameterized phenomenologically
by the form factor parameters at the BBM vertices.

\section{Broken SU(3) form factors and coupling constants}
\label{sec:su3break}
\subsection{Form Factors}
In this paper we describe the results of the NSC97 models where the
form factors depend on the SU(3)$_F$ assignment of the mesons,
rather than on the SU(3)$_F$-irrep structure of the $BB$ channel.
The latter was done for the NSC89 model~\cite{Mae89}. In principle,
we can introduce different form factor masses $\Lambda_{8}$ and
$\Lambda_{1}$ for the $\{8\}$ and $\{1\}$ members of each meson nonet.
However, for practical reasons, we neglect the finer details of the
isoscalar octet and singlet meson mixing, and assign $\Lambda_{1}$
to the physical isoscalar mesons and $\Lambda_{8}$ to the physical
octet mesons. At this stage we are not yet trying to limit the number of
free parameters to an absolute minimum, and so here we also introduce
a separate parameter $\Lambda_K$ for the strange mesons. For example,
for the pseudoscalar mesons we have the following cut-off parameters:
$\Lambda_{1}$ for the $BB\eta'$ vertices, $\Lambda_{8}$ for the $BB\pi$
and $BB\eta$ vertices, and $\Lambda_K$ for the $BBK$ vertices.

\subsection{BBM coupling constants}
For the flavor-symmetry breaking of the coupling constants we use
the $^{3}P_{0}$ mechanism~\cite{Mic69,Rijxx} for the meson-baryon-baryon
coupling. In the $^{3}P_{0}$ model, which is rather successful for
meson decay~\cite{LeY73}, the BBM coupling is due to the rearrangement
of the quark of a virtual quark-antiquark pair in the vacuum and a
valence quark in the baryon. Such a rearrangement leads the initial
baryon state into the final baryon-meson state. The amplitude for the
formation of a meson is calculated from the overlap between the
wave functions of the incoming baryon, the outgoing baryon, the
outgoing meson, and the $q\bar{q}$-pair wave function. For reasons of
simplicity it is usually assumed that the momentum distribution of the
created pair is independent of the momenta.

In scattering, one has to describe not only the emission of mesons, but
also the absorption of mesons. In a Feynman graph a single vertex
implicitly contains both processes and there is no distinction between
emission and absorption. Consider now the $\Lambda N\!K$ vertex as a
specific example. In the quark model the emission of a $K$ is described
by the creation of a non-strange $q\bar{q}$ pair, whereas the absorption
of a $K$ is described by the annihilation of an $s\bar{s}$ pair.
To implement SU(3)$_F$-symmetry breaking within the context of the
$^{3}P_{0}$ model, the usual $^{3}P_{0}$ interaction for decay has to
be generalized. In \cite{Rijxx} this is done by introducing a factor
which describes the transition of a quark from within a baryon to a
quark within a meson, or vice versa. This symmetric treatment of the
``moving'' quarks and the pair quarks then leads to a covariant vertex.
Therefore, in \cite{Rijxx} the $^{3}P_{0}$ Hamiltonian for the BBM
couplings is taken as follows
\begin{equation}
  H_{I}= \int\!\!d^3x \int\!\!d^3y F(x-y)
   \left[\bar{q}(x)O_{\bar{q}q}q(x)\right]^{(1)} \otimes
   \left[\bar{q}(y)O_{\bar{q}q}q(y)\right]^{(2)},   \label{HI3p0}
\end{equation}
where the quark-field operators are vectors in flavor space, with
components $q_{i}=(u,d,s)$ and $\bar{q}_{i}=(\bar{u},\bar{d},\bar{s})$.
In Eq.~(\ref{HI3p0}) it is understood that the first factor creates or
annihilates a $q\bar{q}$ pair, whereas the second factor ``moves'' a
quark from the baryon into the meson or vice versa.
The operator $O_{\bar{q}q}$ is a matrix in quark-flavor space which
is diagonal if we assume there is no quark mixing. Since in general
it will break SU(3) and SU(2) symmetry, it will be of the form
\begin{equation}
    (O_{\bar{q} q})_{i,j}= \left(\begin{array}{ccc}
                         \gamma_{u} &   0        &     0     \\
                            0       & \gamma_{d} &     0     \\
                            0       &    0       & \gamma_{s}
                          \end{array}\right),    \label{Oqq}
\end{equation}
where the pair-creation constants $\gamma_{u}$, $\gamma_{d}$, and
$\gamma_{s}$ are unequal.

The space-time structure will not play an important role in this paper.
We assume that the effects from the overlap of the wave functions can be
effectively absorbed into the $\gamma$ constants. Hence, our matrix
elements will contain an SU(2)$_S$ part due to the spins, and an
SU(3)$_F$ part due to the flavors, and so from here on we can restrict
ourselves to deal explicitly only with the spin and flavor part of the
interaction Hamiltonian density.

Writing the $O_{ij}$ matrix elements in terms of the SU(3) generators
$F_{i}=\lambda_{i}/2$, $(i=1,\ldots,8)$, where $\lambda_{i}$ are the
well-known Gell-Mann matrices, we have
\begin{equation}
   O_{\bar{q}q}=\gamma_{0}F_{0}+\gamma_{3}F_{3}+\gamma_{8}F_{8},
\end{equation}
with $F_{0}$ the unit matrix. We neglect isospin breaking of the
coupling constants, and we set $\gamma_{u}=\gamma_{d}\equiv\gamma_{n}$.
This gives
\begin{equation}
  \gamma_{0}=\frac{1}{3} (2\gamma_{n}+\gamma_{s}), \ \
  \gamma_{3}=0, \ \
  \gamma_{8}=\frac{2}{\sqrt{3}}(\gamma_{n}-\gamma_{s}).
\end{equation}
For $\gamma_{u}=\gamma_{d}=\gamma_{s}$ one has exact SU(3)$_F$ symmetry,
assuming there is no breaking due to differences between the wave
functions of different quark flavors.
For $\gamma_{u}=\gamma_{d}\neq\gamma_{s}$, one gets a breaking of the
coupling constants. In this case, there is still isospin symmetry,
SU(2)$_I$, but SU(3)$_F$ is broken. As an operator in flavor space,
the interaction (\ref{HI3p0}) can now be written as
\begin{eqnarray}
   {\cal H}_{I} &=& \left[\gamma_{0}F_{0}+\gamma_{8}F_{8}\right]^{(1)}
             \otimes\left[\gamma_{0}F_{0}+\gamma_{8}F_{8}\right]^{(2)}
                                  \nonumber\\
   &=& {\cal H}_{I}^{(1)}+{\cal H}_{I}^{(8)}+
       {\cal H}_{I}^{(8\otimes 8)},
\end{eqnarray}
where the singlet interaction, ${\cal H}_{I}^{(1)}$, and the octet
interaction, ${\cal H}_{I}^{(8)}$, correspond to the $\gamma_0^2$ and
$\gamma_{0}\gamma_{8}$ terms, respectively. Because we expect that the
SU(3)$_F$ symmetry is not broken by more than 20\%, the $8\otimes 8$
interaction as given by the $\gamma_{8}^{2}$ term will be rather small.
In the $^{3}P_{0}$-model calculations~\cite{Rijxx} the $8\otimes 8$
piece is implicitly included and can readily be retrieved from the
results by translating $\gamma_{n}$ and $\gamma_{s}$ into $\gamma_{0}$
and $\gamma_{8}$.

The $^{3}P_{0}$ model has approximately SU(6)$_W$ symmetry~\cite{Bar65}.
Therefore, in \cite{Rijxx} the BBM couplings were evaluated using
SU(6)$_W$ wave functions. Since in SU(6)$_W$ the majority of the mesons
have $W=1$, we use here the results for the SU(3)$_F$ breaking for $W=1$
for all mesons. In terms of the SU(3)-flavor breaking parameter
$\lambda_{\rm fsb}=\gamma_{s}/\gamma_{n}$, the modification to the
pseudovector coupling constants is as follows. For the $K$,
\begin{eqnarray}
   f_{\Lambda N\!K} &\rightarrow& f_{\Lambda N\!K}
                  -f_{\Lambda N\!K}(1-\lambda_{\rm fsb}),       \nonumber\\
   f_{\Lambda\Xi K} &\rightarrow& f_{\Lambda\Xi K}
                  -f_{\Lambda\Xi K}(1-\lambda_{\rm fsb}),       \nonumber\\
   f_{\Sigma N\!K} &\rightarrow& f_{\Sigma N\!K}
                  -f_{\Sigma N\!K}(1-\lambda_{\rm fsb}),        \nonumber\\
   f_{\Sigma\Xi K} &\rightarrow& f_{\Sigma\Xi K}
                  -f_{\Sigma\Xi K}(1-\lambda_{\rm fsb}),
\end{eqnarray}
for the $\eta_8$,
\begin{eqnarray}
   f_{\Sigma\Sigma\eta_8} &\rightarrow& f_{\Sigma\Sigma\eta_8}
                  -{\textstyle\frac{1}{3}}f_{\Sigma\Sigma\eta_8}
                   (1-\lambda_{\rm fsb}^2),                     \nonumber\\
   f_{\Xi\Xi\eta_8} &\rightarrow& f_{\Xi\Xi\eta_8}
                  -{\textstyle\frac{8}{9}}f_{\Xi\Xi\eta_8}
                   (1-\lambda_{\rm fsb}^2),
\end{eqnarray}
and for the $\eta_0$,
\begin{eqnarray}
   f_{\Lambda\Lambda\eta_0} &\rightarrow& f_{\Lambda\Lambda\eta_0}
               -f_{\Lambda\Lambda\eta_0}(1-\lambda_{\rm fsb}^2), \nonumber\\
   f_{\Sigma\Sigma\eta_0} &\rightarrow& f_{\Sigma\Sigma\eta_0}
               +{\textstyle\frac{1}{3}}f_{\Sigma\Sigma\eta_0}
                (1-\lambda_{\rm fsb}^2),                         \nonumber\\
   f_{\Xi\Xi\eta_0} &\rightarrow& f_{\Xi\Xi\eta_0}
               -{\textstyle\frac{4}{3}}f_{\Xi\Xi\eta_0}
                (1-\lambda_{\rm fsb}^2).
\end{eqnarray}
Similar expressions apply for the vector and scalar mesons.

\section{Fit to $\protect\bbox{Y\!N}$ total cross sections}
\label{sec:YNfit}
In principle, the potential model contains four free parameters for
each type of meson exchange, and (at this stage) three cut-off parameters
to regularize the corresponding baryon-baryon-meson vertices.
As mentioned earlier, the advantage of abandoning the SU(3)-irrep
scheme for the cut-off parameters is that now the fit to the $Y\!N$
(and $N\!N$) scattering data fixes all parameters, and so the model
can be readily extended to the strangeness $S=-2$, $-3$, and $-4$
sectors. The SU(3)-irrep scheme requires the introduction of new
cut-off parameters for these channels, whereas there are no experimental
data to fix them.

We have made six different fits to the $Y\!N$ scattering data,
including partial waves up to $L=2$. The data we use are tabulated
in Ref.~\cite{Mae89}, and are at sufficiently low energies that the
contributions of the higher partial waves can be safely neglected.
The $N\!N$ interaction puts constraints on most of the parameters,
and so we are left with only a limited set of parameters that we can
vary. The parameters common to all six models are given in
Table~\ref{tabparcomm}. For the remaining parameters we chose six
fixed values for the magnetic vector-meson $F/(F+D)$ ratio $\alpha_V^m$,
ranging from $\alpha_V^m=0.4447$ to $\alpha_V^m=0.3647$.
Adjusting the scalar mixing angle $\theta_S$ and the SU(3)-flavor
breaking parameters $\lambda_{\rm fsb}$, equally good fits to the
$Y\!N$ scattering data have been obtained.
The fitted parameters are given in Table~\ref{tabparfit}, where the
models NSC97a through NSC97f are classified by their different choices
for the magnetic vector-meson $F/(F+D)$ ratio $\alpha_V^m$.

The aim of the present study is to construct a set of models which
give essentially the same fit to the $Y\!N$ scattering data, but
which differ somewhat in the details of their parameterization.
These models will then be used to study the model dependence in
calculations of hypernuclei and in their predictions for the $S=-2$,
$-3$, and $-4$ sectors. Especially for the latter application, these
models will be the first models for the $S<-1$ sector to have
their theoretical foundation in the $N\!N$ and $Y\!N$ sectors.
The results for the $S<-1$ sector will be presented in a future
publication.

The $\chi^2$ on the 35 $Y\!N$ scattering data for the different
models is given in Table~\ref{tabchimod}. Although there is some
variation in the description of some experiments from one model
to the next, these variations are rather small. The total $\chi^2$
on all data varies only a little, and is found to be 15.68, 15.82,
15.62, 15.76, 16.06, and 16.67, for models NSC97a through NSC97f,
respectively. The capture ratio at rest, given in the last column
of Table~\ref{tabchimod}, is defined as
\begin{equation}
   r_R=\frac{1}{4}\,\frac{\sigma_s(\Sigma^-p\rightarrow\Sigma^0n)}
                    {\sigma_s(\Sigma^-p\rightarrow\Lambda n)
                    +\sigma_s(\Sigma^-p\rightarrow\Sigma^0n)}
      +\frac{3}{4}\,\frac{\sigma_t(\Sigma^-p\rightarrow\Sigma^0n)}
                    {\sigma_t(\Sigma^-p\rightarrow\Lambda n)
                    +\sigma_t(\Sigma^-p\rightarrow\Sigma^0n)},
                                \label{caprest}
\end{equation}
where $\sigma_s$ is the total reaction cross section in the singlet
$^1S_0$ partial wave, and $\sigma_t$ the total reaction cross section
in the triplet-coupled $^3S_1$-$^3D_1$ partial wave, both at zero
momentum. In practice these cross sections are calculated at
$p_{\rm lab}=10$ MeV/$c$, which is close enough to zero.
The capture ratio at nonzero momentum is the capture ratio in flight,
defined as
\begin{equation}
   r_F=\frac{\sigma(\Sigma^-p\rightarrow\Sigma^0n)}
            {\sigma(\Sigma^-p\rightarrow\Lambda n)
            +\sigma(\Sigma^-p\rightarrow\Sigma^0n)}.  \label{capflight}
\end{equation}
This capture ratio turns out to be rather constant up to lab momenta
of about 150 MeV/$c$. Obviously, for very low momenta the cross
sections are almost completely dominated by $S$ waves, and so the
capture ratio in flight converges to the capture ratio at rest.

The comparison to the experimental data for models NSC97a, NSC97c, and
NSC97f is shown in Fig.~\ref{fig:moddat}. Models NSC97b, NSC97d, and
NSC97e give similar results, but are left out to avoid overcrowding
in the figures.
The $\Lambda p$ total cross section in Fig.~\ref{fig:moddat}(b) shows a
pronounced cusp of almost 50 mb at the $\Sigma^+n$ threshold, which is
caused by the coupling of the $\Lambda N$ and $\Sigma N$ channels and
the rather strong interaction in the $^3S_1$-wave $\Sigma N$ channel.
Because the cusp occurs over a very narrow momentum range, it is
hard to see this effect experimentally. Indeed, the old bubble-chamber
data~\cite{Kad71,Hau77} have too large error bars to identify any
possible cusp effect. (Note that these data have not been used in our
fits.)

It should be noted that the $\Sigma^+p$ and $\Sigma^-p$ elastic cross
sections are not the ``true'' total cross sections. The latter are
hard to measure because of the large Coulomb contribution at forward
angles. The cross sections that were measured are defined
as~\cite{Eis71}
\begin{equation}
  \sigma=\frac{2}{\cos\theta_{\rm max}-\cos\theta_{\rm min}}
         \int_{\theta_{\rm min}}^{\theta_{\rm max}}
         \frac{d\sigma(\theta)}{d\cos\theta}d\cos\theta,
\end{equation}
with typical values $-0.2$ to $-0.5$ for $\cos\theta_{\rm min}$ and
$0.3$ to $0.5$ for $\cos\theta_{\rm max}$. In order to stay as close
as possible to the plotted experimental data, the theoretical curves
in Figs.~\ref{fig:moddat}(c) and (d) have been calculated with
$\cos\theta_{\rm min}=-0.5$ and $\cos\theta_{\rm max}=0.5$.
The Heidelberg group~\cite{Eis71} also presents elastic differential
cross sections for $\Sigma^{\pm}p$ scattering at $p_{\Sigma^{\pm}}=170$
and 160 MeV/$c$, respectively. The corresponding potential model
predictions are plotted in Fig.~\ref{fig:Eisdsg}; again, only models
NSC97a, NSC97c, and NSC97f are shown.

Although the six models give an equally good description of the (few)
$Y\!N$ scattering data, the different choices for $\alpha_V^m$ give
rise to different properties on a more detailed level. This implies
that these scattering data do not unambiguously determine the
$Y\!N$ interaction. For example, in Fig.~\ref{fig:Sphs} we show the
wide spread in the $\Lambda p$ $^1S_0$ and $\Sigma^+p$ $^3S_1$ phase
shifts which, according to the results in Table~\ref{tabchimod}, are
still compatible with the scattering data.
Also, the $S$-wave scattering lengths in the four $Y\!N$ channels
exhibit a fair amount of variation from one model to the next, as shown
in Table~\ref{tabmodscat}.

As will be discussed in the next section, the differences among these
models in applications other than low-energy $Y\!N$ scattering are even
more pronounced. As a consequence, they will provide important
information to further pin down the $Y\!N$ interaction.
It is found that especially NSC97f exhibits nice features when applied
to hypernuclear systems. Therefore, rather than providing many tables
with results for all the models, we will here only give some results
for NSC97f. The phase shifts for $\Sigma^+p$ and $\Lambda p$ scattering
are given in Tables~\ref{tabphs:Sp} and \ref{tabphs:Lp}, respectively.
Predictions for the total cross sections in the $\Lambda p$ channel
above the $\Sigma N$ thresholds are given in Table~\ref{tabsgt:Lp},
while those for the total nuclear (i.e., without Coulomb) cross sections
in the $\Sigma^-p$ channel are given in Table~\ref{tabsgt:Sp}.

\section{G-matrix analyses of NSC97 models}
\label{sec:hypnuc}
The properties of hypernuclear systems are linked closely to the
underlying $Y\!N$ interactions. Since the free-space $Y\!N$ scattering
data are sparse at the present stage, it is quite important to test
our OBE models through the study of hypernuclear phenomena.
Especially, the coming precise data of $\gamma$-ray observation from
$\Lambda$ hypernuclei will provide very valuable information on the
spin-dependent forces such as spin-spin and spin-orbit interactions.
Effective $Y\!N$ interactions in a nuclear medium, which reflect
the properties of the bare interactions, can be derived using the
G-matrix procedure. One of the authors (Y.Y.) and his collaborators
performed the G-matrix calculations in nuclear matter with the various
OBE models by the Nijmegen~\cite{Nag77,Nag79,Mae89} and J\"ulich
groups~\cite{Hol89,Reu94}, and found specific differences among
them~\cite{Yam85,Yam90,Yam94}.
Here, we discuss the properties of G-matrix interactions derived from
the NSC97 models in comparison with the old NSC89 version~\cite{Mae89},
and the hard-core models D~\cite{Nag77} and F~\cite{Nag79} (referred
to as NHC-D and NHC-F, respectively). In order to compare the present
results with the past works~\cite{Yam85,Yam90,Yam94}, the calculations
are done in the same framework:
We adopt the simple QTQ prescription for the intermediate-state
spectrum, which means that no potential term is taken into account
in the off-shell propagation. As discussed later, this procedure is
reliable enough to investigate the feature of spin-dependent terms.

In Table~\ref{tabgmat1} we show the potential energies $U_\Lambda$
for a zero-momentum $\Lambda$ and their partial-wave contributions
$U_\Lambda(^{2S+1}L_J)$ at normal density ($k_F$=1.35 fm$^{-1}$) for
the NSC97 models, where a statistical factor $(2J+1)$ is included in
$U_\Lambda(^{2S+1}L_J)$. It is seen that the values for each state vary
smoothly from NSC97a to NSC97f. The obtained values for $U_\Lambda$
are not so far from the well depths ($\sim$28 MeV) of $\Lambda$-nucleus
Woods-Saxon potentials as obtained from analyses of the
$(\pi^{+},K^{+})$ reaction data~\cite{Mil88,Pil91,Has96},
though the comparison should be only considered on a qualitative level.
It should be noted here that the odd-state interactions, which are
uncertain experimentally, are very different among the various OBE
models. In the case of the NSC97 models, the odd-state contributions
are found to be strongly repulsive. On the other hand, they are strongly
attractive, weakly attractive, and almost vanishing in the case of
NHC-D, NHC-F, and NSC89, respectively~\cite{Yam94}.
The stronger odd-state repulsion of the NSC97 models is compensated
by the also stronger even-state attraction.

It is noted that the relative ratios of $U_\Lambda(^1S_0)$ and
$U_\Lambda(^3S_1)$ are very different among the NSC97 models, as seen
in Table~\ref{tabgmat1}, indicating different spin-spin interactions.
In order to see the spin-dependent features of the G-matrix interactions
more clearly, we obtain the contributions to $U_\Lambda$ from the
spin-independent, spin-spin, $LS$, and tensor components of
the G matrices, denoted as $U_0$, $U_{\sigma\sigma}$, $U_{LS}$,
and $U_{T}$, respectively, by the following transformations:
\begin{eqnarray}
  && U_0(S) = \frac{1}{4}\left\{U(^3S_1)+U(^1S_0)\right\}, \nonumber\\
  && U_{\sigma\sigma}(S) = \frac{1}{12}\left\{U(^3S_1)-3 U(^1S_0)\right\},
                                                           \nonumber\\
  && U_0(P) = \frac{1}{12}\left\{U(^3P_0)+U(^3P_1)+U(^3P_2)
                            +3U(^1P_1)\right\},            \nonumber\\
  && U_{\sigma\sigma}(P) = \frac{1}{36}\left\{U(^3P_0)+U(^3P_1)+U(^3P_2)
                            -9U(^1P_1)\right\},            \nonumber\\
  && U_{LS}(P) = \frac{1}{12}\left\{-2U(^3P_0)-U(^3P_1)+U(^3P_2)\right\},
                                                           \nonumber\\
  && U_{T}(P) = \frac{1}{72}\left\{-10U(^3P_0)+5U(^3P_1)-U(^3P_2)\right\}.
                  \label{eq:trans}
\end{eqnarray}
The obtained values are shown in Table~\ref{tabgmat2}, where
also the ones for NSC89, NHC-D, and NHC-F are given for comparison.
We can see here the nice correlation between the $\alpha_V^m$
values taken in the NSC models and the strengths of the spin-spin
interactions in even states; the smaller value of $\alpha_V^m$ leads
to the more repulsive strength. This marked difference of the spin-spin
interactions for NSC97a--f will show up characteristically in hypernuclear
spectra, which should be tested in comparison with experimental data.
On the other hand, the differences of $LS$ components amongst the
Nijmegen models turn out to be less remarkable than the spin-spin ones.
It is notable here that also the strengths of the $LS$ interactions
vary smoothly with the $\alpha_V^m$ values in the NSC97 models.
The detailed discussion of spin-orbit components is given later.

As is well known, there remains some ambiguity in the lowest-order
G-matrix approximation concerning the intermediate spectrum in the
propagator. The choice of a continuous intermediate-energy spectrum
(CIES), extended smoothly from the on-shell one, leads to the
considerable gain of $U_\Lambda$ values in comparison with the QTQ
prescription~\cite{Yam90,Yam94,Sch95}.
Now, let us demonstrate that the spin-dependent parts are not so much
affected by these different treatments, in spite of the considerable
change of the spin-independent part $U_0$.
In Table~\ref{tabgmat3} the above-defined components for NSC97f
are compared for the cases of QTQ and CIES, where the results are
given at $k_F=1.35$ and 1.0 fm$^{-1}$.
It should be noted here that the differences of the spin-dependent
parts are far smaller than those of the spin-independent ones $U_0$,
especially in the case of $k_F=1.0$ fm$^{-1}$.
The reason is that these spin-dependent contributions are determined
essentially by the differences of the partial-wave contributions, and
the induced changes cancel out considerably.
In addition, the $P$-state contributions are far less sensitive
to the treatment of the intermediate spectrum than the $S$-state ones.
Due to the same reasons, the density dependencies of the
spin-dependent parts are considerably weaker than those of the
spin-independent parts.

If the nucleon rearrangement effect is taken into account~\cite{Yam85},
the values of $U_\Lambda$ are multiplied by $(1-\kappa_N)$,
$\kappa_N=0.10 \sim 0.15$ (at normal density) being the
average correlation probability for nucleons.
In Table~\ref{tabgmat3} we find that the QTQ results without
the $(1-\kappa_N)$ correction simulate roughly the CIES results
with this correction.

Thus, we can say that the G-matrix interactions are reliable enough for
bridging the spin-dependent terms of the OBE models with hypernuclear
spectra separately from the ambiguities of the spin-independent parts
alluded to above. A convenient approach is, for instance, to adjust
the spin-independent parts adequately so as to reproduce the
experimental $\Lambda$ binding energies in applying the G-matrix
interactions to structure calculations of hypernuclei~\cite{Yam94}.

Let us discuss the $\Lambda$ $\ell$-$s$ potentials in hypernuclei,
which are derived from the $LS$ and anti-symmetric $LS$ ($ALS$)
components of our G-matrix interactions, in comparison with the
corresponding nucleon one.
In the Scheerbaum-approximation~\cite{Sch76} the $\ell$-$s$
potential is related to the two-body $LS$ $(ALS)$
interaction as follows:
\begin{eqnarray}
   && U^{ls}_B(r) = K_B\, {1 \over r} \, {d\rho \over dr}\,
      {\bf l}\cdot{\bf s} \qquad {\rm with} \quad B=N ,\Lambda, \nonumber\\
   && K_N = -{\pi \over 2} S_{LS} \quad {\rm and} \quad K_\Lambda=
            -{\pi \over 3} (S_{LS}+S_{ALS}),                    \nonumber\\
   && S_{LS,ALS} = {3 \over \bar q} \int^\infty_0 r^3
                   j_1(\bar q r)\, G_{LS,ALS}(r)\, dr,
             \label{eq:so}
\end{eqnarray}
where $G_{LS}(r)$ and $G_{ALS}(r)$ are the $LS$ and $ALS$ parts of the
G-matrix interactions in configuration space, respectively, and $\rho(r)$
is the nuclear density distribution. We take here $\bar q=0.7$ fm$^{-1}$
simply in the same way as~\cite{Sch76}, since the results are
insensitive to the value of $\bar q$.

Table~\ref{tabgmat4} shows the values of $K_\Lambda$ and $S_{LS,ALS}$
obtained from the $LS$ and $ALS$ parts of the $\Lambda N$ G-matrix
interactions derived from various Nijmegen models, where the
G matrices are calculated at $k_F=1.0$ fm$^{-1}$. Due to the reason
mentioned above, the $LS$ and $ALS$ parts are very insensitive to $k_F$.
For comparison we give also $K_N$ and $S_{LS}$ for a typical $N\!N$
G-matrix interaction ($G0$)~\cite{Spr72} derived from the Reid soft-core
potential. Here, it should be noted that the effective strengths
$S_{LS}$ of the $\Lambda N$ $LS$ interactions are not so small compared
to that of $N\!N$. In the case of NSC97f, for instance, the absolute
value of $S_{LS}$ is smaller than that of $N\!N$ by 66\%, but
the $K_\Lambda$ value is smaller than the $K_N$ one by 32\%.
There are two reasons why the values of $K_\Lambda$ become so small
compared to that of $K_N$. One is that $K_\Lambda$ is smaller than
$K_N$ kinematically by 2/3, which is determined by the ratio of
the number of $N\!N$ and $\Lambda N$ $^3$O-bonds in the nucleus.
The other reason is that the $LS$ contribution is canceled substantially
by the $ALS$ one in the $\Lambda N$ case.

Furthermore, be aware that the ratio of the $\Lambda$ and $N$
$\ell$-$s$ splitting energies should be further reduced compared with
$K_\Lambda/K_N$. First, the value of $K_N$ obtained with $G0$ accounts
for only about $60 \sim 70 \%$ of the empirical $\ell$-$s$ splitting;
the additional contributions are supposed to come from the many-body
correlations related to the Pauli exclusion effect~\cite{And81}, which
are not expected to be present for a $\Lambda$ particle.
Second, the $\Lambda$ single-particle wave function should extend
farther than the $N$ one due to its smaller binding energy,
which leads to a reduction of the $\Lambda N$ matrix elements.
In the case of $^{17}_\Lambda$O, for instance, it was found that the
$\Lambda$ $\ell$-$s$ splitting energy evaluated with the precise
$\Lambda$ wave function is reduced by $\sim 25 \%$ from that with
the single harmonic oscillator one~\cite{Dal97}. Then, the ratio of
$\Lambda$ and $N$ $\ell$-$s$ splitting energy is estimated as about
one half of $K_\Lambda /K_N$. Precise calculations of $\Lambda$
splitting energies can be done with use of the above values of
$K_\Lambda$ or $G_{LS,ALS}(r)$ themselves~\cite{Dal97}.

Thus, it is concluded that $\Lambda$ $\ell$-$s$ splitting energies
in hypernuclei are likely to be very small compared to nucleon ones,
even if the $\Lambda N$ $LS$ interaction is not so much weaker than
the $N\!N$ one. Precise measurements of $\Lambda$ $\ell$-$s$ splitting
energies are crucially important to extract information on the
two-body $LS$ and $ALS$ interactions.

Finally, we comment on the properties of the $\Sigma N$ G-matrix
interactions.
The calculations are done in the same way as in Ref.~\cite{Yam94}.
The QTQ spectra are adopted in $\Sigma N$ intermediate states, but
continuous intermediate ones are taken into account in $\Lambda N$
states coupled to $\Sigma N$ channels. The potentials in the
intermediate $\Lambda N$ states only slightly influence the real
parts of the $\Sigma N$ G matrices, but work decisively on their
imaginary parts related to the conversion width in nuclear matter.
The reason is that the imaginary part is determined by the
energy-conserving transition from the starting $\Sigma N$ state
to the $\Lambda N$ one.
We calculate here $\Sigma$ single-particle potentials $U_\Sigma$
and conversion widths $\Gamma_\Sigma$ for the NSC97 models.
The obtained results are more or less similar to each other.
In Table~\ref{tabgmat5} the calculated values of $U_\Sigma$
and $\Gamma_\Sigma$ at $k_F=1.0$ fm$^{-1}$ for the NSC97e and NSC97f
are compared with those for the other Nijmegen models.
We find remarkable differences among the models.

The $\Gamma_\Sigma$ directly reflects the strength of the
$\Sigma N$-$\Lambda N$ coupling interaction, and those of the
NSC97 models turn out to be considerably smaller than for NSC89.
It is worthwhile to say that the moderate $\Sigma N$-$\Lambda N$
coupling interactions of the NSC97 models are free from
possible troubles which appear in applications of the NSC89 model
to hypernuclear systems due to its too strong $\Lambda N$-$\Sigma N$
coupling.

Recently, the existence of $^4_\Sigma$He has been
confirmed~\cite{Nag98}, which gives valuable information on
the $\Sigma N$ interaction. The observed values of the $\Sigma$
binding energy, $B_\Sigma$, and the width are $4.4 \pm 0.3 \pm 1$ MeV
and $7.7 \pm 0.7^{+1.2}_{-0.0}$ MeV, respectively.
As discussed in Ref.~\cite{Har90}, the strong spin-isospin
dependence shows up in the $\Sigma N$ interaction, and
the value of $B_\Sigma$ is determined mainly by the attractions
in the $T=1/2$ $^3S_1$ and $T=3/2$ $^1S_0$ states.
It should be noted that the NSC97 models are adequately attractive
in these states, as well as the other Nijmegen models.
Be careful that our calculated values of $\Gamma_\Sigma$ should
not be compared directly to the above experimental one.
Because of weak binding of the $\Sigma$, the wave function
extends outwards and is of small overlap with the nucleon ones,
which leads likely to a remarkable reduction of $\Gamma_\Sigma$.
It is an open problem to perform exact four-body calculations
on the basis of these OBE models.

\section{Discussion and Outlook}
\label{sec:conc}
The NSC97 models are an important step forward in the realization
of a program where the baryon-baryon interactions for scattering and
hypernuclei can be described in the context of broken SU(3)$_F$ symmetry.

First, it turns out that starting from the soft-core OBE model for
$N\!N$, we are indeed able to achieve a very good description of the
$Y\!N$ data and at the same time maintain values for the free parameters
which are consistent with the present view on low-energy hadron physics.
For example, the value used for the $F/(F+D)$ ratio $\alpha_{PV}$ for
the pseudoscalar mesons is the same as that found in the weak interactions;
see, e.g., Ref.~\cite{Dum83}.
Also, the range of values used for the magnetic ratio of the vector
mesons is compatible with the estimates from static and non-static
SU(6)~\cite{Sak65}.

Second, for the first time the soft-core model NSC97f passes the tests
from the hypernuclear studies very satisfactorily.
It is no longer necessary to introduce a phenomenological spin-spin
interaction for the $\Lambda N$ systems, as was the case for the
NSC89 model~\cite{Mae89}, see Ref.~\cite{Yam90}. This is an important
achievement with the NSC97 models.

Third, the NSC97 models give parameter-free predictions for the
$S=-2,-3,-4$ two-body systems.
In the $S=-2$ systems, the experimental information is limited to the
ground states of $_{\Lambda\Lambda}^{\,\ 6}$He,
$_{\Lambda\Lambda}^{\;10}$Be, and $_{\Lambda\Lambda}^{\;13}$B, from which
it is inferred that $\Delta B_{\Lambda\Lambda}=4-5$MeV, corresponding
to a rather strong attractive $\Lambda\Lambda$ interaction.
The estimate for the $^{1}S_{0}$ $\Lambda\Lambda$ matrix element in
$_{\Lambda\Lambda}^{\,\ 6}$He for NHC-D~\cite{Nag77} is
$\Delta B_{\Lambda\Lambda}=4$ MeV, in agreement with the experimental
observation.
For more details we refer to Ref.~\cite{Yam94}. Now, the characteristic
feature of NHC-D is that, instead of a scalar nonet, there is only a
scalar singlet. This makes the scalar central attraction independent
of the baryon-baryon channel, and hence equally strong as in $N\!N$.
However, in the soft-core models constructed sofar, we have nearly
ideal mixing for $q\bar{q}$ states, which implies that
\[
 |V_{\Lambda\Lambda}(0^{+})| < |V_{\Lambda N}(0^{+})| < |V_{N\!N}(0^{+})|,
\]
which leads to much weaker attractive potentials than in the case of
NHC-D in the $\Lambda\Lambda$ and $\Xi N$ systems. For example,
an estimate for the $\Lambda\Lambda(^{1}S_{0})$ scattering length,
based on $\Delta B_{\Lambda\Lambda}$ quoted above, is
$a_{\Lambda\Lambda}(^{1}S_{0}) \approx -2.0$ fm~\cite{Tan65,Bod65}.
In the NSC97 models we obtain values between --0.3 and --0.5 fm.
The only way to produce stronger $\Lambda\Lambda$ forces is to go to
smaller $\theta_S$ and ipso facto a smaller $\alpha_{S}$.
However, when we tried this for the soft-core OBE models, we
produced a $\Lambda N(^{1}S_{0})$ bound state.
On the other hand, preliminary results from a potential model which
includes also the two-meson-exchange contributions within the present
framework, do show the apparently required attraction in the
$\Lambda\Lambda$ interaction. This model is currently under further
development.

Finally, to put the NSC97 models in perspective, we conclude by
discussing the present situation of the Nijmegen models for the
central, spin-spin, and spin-orbit interactions with respect to
information from hypernuclear studies.

\subsubsection{Central interaction}
The $\Lambda$ well depth $U_{\Lambda}$ in nuclear medium is of basic
importance in hypernuclear physics. The data of the middle and heavy
$\Lambda$ hypernuclei at BNL~\cite{Pil91} and KEK~\cite{Has96} play an
essential role, because it seems rather ambiguous to extrapolate
$U_{\Lambda}$ from $\Lambda$ binding energies in light systems.
The phenomenological analyses that have been performed for experimental
$B_\Lambda$ values with the use of Woods-Saxon potentials, indicate
a depth of $\sim 28$ MeV~\cite{Mil88,Has96}.
Some OBE models, including the present NSC97 ones, reproduce this
value fairly well in the lowest-order G-matrix theory.
{}From a fundamental many-body point of view, however, the comparison
should be considered as being qualitative, because of ambiguities
in this G-matrix approximation, especially in the spin-independent parts.
One of the features of NSC97 models is that the odd-state interactions
are strongly repulsive, but are compensated by the strong even-state
attractions. This in contrast to the earlier Nijmegen models.
It is an open problem to test this feature, for instance, by
analyzing the experimental $\Lambda$ single-particle spectra
in medium and heavy hypernuclei~\cite{Yam94}.

\subsubsection{Spin-spin interaction}
The spin-doublet splittings ($J_{>,<}=J_c \pm s^\Lambda_{1/2}$) of
several hypernuclei have been analyzed extensively by Yamamoto
{\it et al.}~\cite{Yam94} using the G-matrix interactions derived
from the Nijmegen and J\"ulich potentials.
As seen in Table~\ref{tabgmat2}, the strengths of the spin-spin
interactions are very different among the Nijmegen models,
where the most repulsive (attractive) is that of NSC89 (NHC-D).
Those of the J\"ulich potentials are known to be more attractive
than NHC-D~\cite{Yam94}. The spin-spin interactions show up in the
differences of the $^1S_0$ and $^3S_1$ phase shifts.
The values obtained for $\Lambda p$ scattering at $p_\Lambda=200$ MeV/c
are $-18.89^{\circ}$, $-15.33^{\circ}$, $-10.55^{\circ}$, $-3.34^{\circ}$,
$1.40^{\circ}$, $5.60^{\circ}$, $9.14^{\circ}$, $-4.17^{\circ}$, and
$2.02^{\circ}$ for NSC97 models a, b, c, d, e, f, NSC89, NHC-D, and
NHC-F, respectively. Here, positive (negative) values mean repulsive
(attractive) spin-spin interactions.
Comparing these values to those for $U_{\sigma\sigma}$, we find a nice
systematic correspondence between them. The experimental manifestation
of the $\Lambda N$ spin-spin interaction is found in the $0^+$-$1^+$
doublet states of $^4_\Lambda$H and $^4_\Lambda$He~\cite{Bed79}, where
the $J_< =0^+$ state is below the $J_> =1^+$ one by about 1 MeV.
The analysis of $^4_\Lambda$H with the G-matrix interactions
indicates that the spin-spin interaction should be repulsive and
its adequate strength is between those of NHC-F and NSC89~\cite{Yam94}.
Then, that of NSC97e or NSC97f seem to be of adequate strength,
though a definite conclusion should be based on more elaborate
four-body calculations. A complementary indication can be obtained
from the exact three-body calculations of $^3_\Lambda$H by Miyagawa
{\it et al.}~\cite{Miy93,Miy95}, where the repulsive (attractive)
spin-spin interaction of NSC89 (J\"ulich A) is shown to be adequate
(inadequate) to reproduce the experimental $\Lambda$ binding energy.
Recently, Miyagawa performed the same calculations using the NSC97
models~\cite{Miy98}: The model NSC97f, whose spin-spin interaction is
of the most repulsive among the NSC97 models, reproduces a reasonable
$\Lambda$ binding energy. On the other hand, the model NSC97e gives
rise to only a very weakly bound state compared to the experimental
one, and no bound states are obtained for models NSC97a--d.
Thus, the $^3_\Lambda$H problem turns out to be one of the critical
tests for the spin-spin interactions.
Results for the NSC97 potentials with regard to the Carlson-Gibson
computation~\cite{Car91} of the $_{\Lambda}^{3}$He, $_{\Lambda}^{4}$He,
and $_{\Lambda}^{5}$He hypernuclei are not available yet.

The ground-state doublet splitting energies of some light $p$-shell
hypernuclei are also indicative of the spin-spin interactions.
The shell-model analyses of $^{10}_{\ \Lambda}$B, $^{11}_{\,\,\Lambda}$B,
$^{12}_{\ \Lambda}$C, and $^{12}_{\ \Lambda}$B with the G-matrix
interactions showed that the repulsive spin-spin interactions such as
NHC-F and NSC89 make the $J_<$ states lower than the $J_>$
states~\cite{Yam94}. (Experimentally the ground-state spins of
$^{11}_{\,\,\Lambda}$B and $^{12}_{\ \Lambda}$B are $J_< =5/2^+$ and
$1^-$, respectively.)
This situation is altered by the $LS$ and $ALS$ interactions, however,
which works more attractively on the $J_>$ states against the spin-spin
interaction. For instance, the spin-spin interaction of NHC-F is
weakly repulsive and makes $J_<$ states slightly lower than $J_>$
states, but this order is reversed by adding the $LS$ and $ALS$
terms~\cite{Yam94}. On the other hand, the spin-spin interaction of
NSC89 is so repulsive that the $J_<$ states are kept lower~\cite{Yam94}.
Although the spin-spin interaction of NSC97f is less repulsive than
that of NSC89, the $J_<$ states are also kept lower, in spite of
adding the $LS$ and $ALS$ ones~\cite{Mot98}.
Considering that the spin-spin interaction of NSC89 is suggested to
be too repulsive~\cite{Yam94}, that of NSC97f is expected to be of
reasonable strength. The less repulsive one of NSC97e is maybe
of lower limit.
Of course, there still remain ambiguities because the strengths of
$LS$ and $ALS$ interactions are not established experimentally.

As new experiments are planned using hypernuclear $\gamma$-ray
spectrometers with the germanium detectors~\cite{Tam98}, there are
good prospects for progress in this sector. For instance, the planned
experiment of the ground-state doublet splitting of $^7_\Lambda$Li
is very promising, because this splitting is considered to be fairly
free from the $LS$ and $ALS$ interactions~\cite{Hiy98}.
In contrast, the $[^8$Be($2^+) \otimes (s_{1/2})_\Lambda]_{5/2^+,3/2^+}$
splitting in $^9_\Lambda$Be is almost purely determined by the $LS$
and $ALS$ interactions~\cite{Hiy98}.
In view of these developments, one can envisage that the $\Lambda N$
spin-spin and spin-orbit interactions will be established rather well
in the coming years.

\subsubsection{Spin-orbit interaction}
The $\Lambda$ $\ell$-$s$ splitting energies in hypernuclei are related
intimately to the two-body $LS$ and $ALS$ components of $\Lambda N$
interactions. It has been observed that the $\Lambda$ $\ell$-$s$
splitting energies are far smaller than the nucleon ones. The first
indication was given by the $^{16}$O$(K^-,\pi^-)^{16}_{\ \Lambda}$O
experiment at CERN~\cite{Bru78}. The splitting of the observed two peaks
of the $[(p_{3/2}^{-1})_n (p_{3/2})_\Lambda]_{0^+}$ and
$[(p_{1/2}^{-1})_n (p_{1/2})_\Lambda]_{0^+}$ configurations was
almost the same as that of the neutron $p_{1/2}$ and $p_{3/2}$
hole states in $^{15}$O, and the splitting of $p$-state $\Lambda$
was estimated to be less than 0.3 MeV.
In the $^{13}$C$(K^-,\pi^-)^{13}_{\ \Lambda}$C experiment at
BNL~\cite{May81}, the $\Lambda$ splitting energy in $^{13}_{\ \Lambda}$C
was obtained as 0.36$\pm$0.3 MeV with help of some theoretical
consideration on the dominant configurations of the peak.
The $^9$Be$(K^-,\pi^-\gamma)^9_\Lambda$Be experiment at BNL~\cite{May83}
also indicates the small $\Lambda$ $\ell$-$s$ splitting.
Only one observed $\gamma$-ray peak suggests that the excited
doublets $[^9$Be$(2^+) \otimes (s_{1/2})_\Lambda]_{3/2,5/2}$
are almost degenerate, where the splitting energy has to be less than
the experimental resolution of 0.1 MeV.
Anyway, the data of $\Lambda$ $\ell$-$s$ splitting energies are yet
still far from a quantitative determination.

In Table~\ref{tabgmat4} the values of $S_{LS,ALS}$ and $K_\Lambda$ for
the Nijmegen models are compared to the corresponding ones of nucleons.
As stressed in the previous section, the $\Lambda N$ $LS$ interaction
is not so small compared with the $N\!N$ one, which seemingly is
contradictory to the above experimental indications.
However, the $\Lambda$ $\ell$-$s$ splitting is likely to be far
smaller than the $N$ one due to the reasons mentioned in the previous
section. Additionally, the coupling effects with core-excited states
also possibly influence the $\Lambda$ $\ell$-$s$ splitting energies.
Recently, Dalitz {\it et al.}~\cite{Dal97} analyzed the excited doublet
states of $^{16}_{\ \Lambda}$O, whose dominant components are
$[(p_{1/2})_N^{-1} (p_{1/2,3/2})_\Lambda]_{0^+,2^+}$. This splitting
energy was shown to be understood on the basis of the $LS$ and $ALS$
terms of G-matrix interactions derived from the Nijmegen models,
if the coupling to core-excited states with a $(s_{1/2})_\Lambda$
are taken into account.
The new experiment at BNL (E929) is now in progress to determine
the $\Lambda$ $\ell$-$s$ splitting in $^{13}_{\ \Lambda}$C by
detecting the $\gamma$-rays from $(p_{3/2})_\Lambda$ and
$(p_{1/2})_\Lambda$ states.
In order to extract information on the underlying $\Lambda N$
$LS$ interaction from the coming data, it will be necessary to
perform an elaborate structure calculation in which core-excited
states are fully taken into account~\cite{Hiy98}.

The spin-orbit interaction is also very interesting from the point of
view of the quark model. Namely, the $P$-wave baryons are hard to
describe by the theory if one keeps the full Fermi-Breit spin-orbit
interaction from gluon exchange~\cite{Isg78}.
For the literature since 1980, see Valcarce {\it et al.}~\cite{Val95}.
Here one finds an indication that meson-exchange between quarks
($\pi,\epsilon,\rho,\omega$, etc.) is a possible solution.
Another possibility is that the inclusion of the decay channels
will be a way out of this problem~\cite{Fuj93}.

\acknowledgments
Part of this work was done while the first author stayed at the INS,
University of Tokyo. We (Th.A.R. and Y.Y) are grateful to
Y.\ Akaishi, T.\ Harada, and T.\ Motoba for many
stimulating discussions and generous hospitality.
V.G.J. Stoks likes to thank T.-S.H.\ Lee and T.\ Kuo for their continued
interest and constructive comments.
We (Th.A.R. and V.G.J.S.) thank J.J. de Swart and R.G.E. Timmermans
for discussions, concerning in particularly the scalar meson properties.
We are also grateful to K.\ Miyagawa for performing the three-body
calculations for the hypertriton with the NSC97 models.
The work of V.G.J.S.\ was supported in part by the U.S.\ Department
of Energy, Nuclear Physics Division, under Contract No.\ W-31-109-ENG-38.

\appendix
\section*{Potential in configuration space}
(a) Pseudoscalar-meson exchange (pseudovector coupling):
\begin{equation}
   V_{PV}(r)=\frac{m}{4\pi}\left[f_{13}^Pf_{24}^P
             \left(\frac{m}{m_{\pi}}\right)^2
             \left[{\textstyle\frac{1}{3}}
                   (\bbox{\sigma}_1\!\cdot\!\bbox{\sigma}_2)\phi_C^1
                   +S_{12}\phi_T^0\right]\right].  \label{Vpseur}
\end{equation}

(b) Vector-meson exchange:
\begin{eqnarray}
  V_V(r)=\frac{m}{4\pi}\Biggl[ && \left\{g_{13}^Vg_{24}^V
         \left[\phi_C^0+\frac{m^2}{2M_{13}M_{24}}\phi_C^1
              -\frac{3}{4M_{13}M_{24}}
              (\Delta\phi_C^0+\phi_C^0\Delta)\right]\right. \nonumber\\
    && \ \ \left.+\left[g_{13}^Vf_{24}^V\frac{m^2}{4{\cal M}M_{24}}
           +f_{13}^Vg_{24}^V\frac{m^2}{4{\cal M}M_{13}}\right]\phi_C^1
           + f_{13}^Vf_{24}^V\frac{m^4}{16{\cal M}^2M_{13}M_{24}}
               \phi_C^2\right\}                             \nonumber\\
    && \ \ +\frac{m^2}{4M_{13}M_{24}}\left\{
            \left(g_{13}^V+f_{13}^V\frac{M_{13}}{{\cal M}}\right)
            \left(g_{24}^V+f_{24}^V\frac{M_{24}}{{\cal M}}\right)\phi_C^1
           +f_{13}^Vf_{24}^V\frac{m^2}{8{\cal M}^2}\phi_C^2\right\}
           {\textstyle\frac{2}{3}}
           (\bbox{\sigma}_1\!\cdot\!\bbox{\sigma}_2)       \nonumber\\
    && \ \ -\frac{m^2}{4M_{13}M_{24}}\left\{
            \left(g_{13}^V+f_{13}^V\frac{M_{13}}{{\cal M}}\right)
            \left(g_{24}^V+f_{24}^V\frac{M_{24}}{{\cal M}}\right)\phi_T^0
           +f_{13}^Vf_{24}^V\frac{m^2}{8{\cal M}^2}\phi_T^1\right\}S_{12}
                                                           \nonumber\\
    && \ \ -\frac{m^2}{M_{13}M_{24}}\left\{\left[
            {\textstyle\frac{3}{2}}g_{13}^Vg_{24}^V
            +g_{13}^Vf_{24}^V\frac{M_{24}}{{\cal M}}
            +f_{13}^Vg_{24}^V\frac{M_{13}}{{\cal M}}\right]\phi_{SO}^0
           +{\textstyle\frac{3}{8}}f_{13}^Vf_{24}^V
             \frac{m^2}{{\cal M}^2}\phi_{SO}^1\right\}{\bf L\!\cdot\!S}
                                                           \nonumber\\
    && \ \ +\frac{m^4}{16M_{13}^2M_{24}^2}\left\{
            g_{13}^Vg_{24}^V+4(g_{13}^Vf_{24}^V+f_{13}^Vg_{24}^V)
            \frac{\sqrt{M_{13}M_{24}}}{{\cal M}}
            +8f_{13}^Vf_{24}^V\frac{M_{13}M_{24}}{{\cal M}^2}\right\}
           \frac{3}{(mr)^2}\phi_T^0Q_{12}                  \nonumber\\
    && \ \ -\frac{m^2}{M_{13}M_{24}}\left\{\left[
             g_{13}^Vg_{24}^V\phi_{SO}^0-f_{13}^Vf_{24}^V
             \frac{m^2}{4{\cal M}^2}\phi_{SO}^1\right]
             \frac{(M_{24}^2-M_{13}^2)}{4M_{13}M_{24}} \right. \nonumber\\
    && \ \ \hspace*{2cm} \left. -(g_{13}^Vf_{24}^V-f_{13}^Vg_{24}^V)
             \frac{\sqrt{M_{13}M_{24}}}{{\cal M}}\phi_{SO}^0\right\}
            {\textstyle\frac{1}{2}}(\bbox{\sigma}_1-\bbox{\sigma}_2)
            \!\cdot\!{\bf L}\Biggr].       \label{Vvec1r}
\end{eqnarray}

(c) Scalar-meson exchange:
\begin{eqnarray}
  V_S(r)=-\frac{m}{4\pi}g_{13}^Sg_{24}^S && \left\{
          \left[\phi_C^0-\frac{m^2}{4M_{13}M_{24}}\phi_C^1\right]
           +\frac{m^2}{2M_{13}M_{24}}\phi_{SO}^0{\bf L\!\cdot\!S}
           +\frac{m^4}{16M_{13}^2M_{24}^2}\,\frac{3}{(mr)^2}
            \phi_T^0Q_{12}           \right.   \nonumber\\
  && \left. +\frac{m^2}{M_{13}M_{24}}\,
             \frac{(M_{24}^2-M_{13}^2)}{4M_{13}M_{24}}\phi_{SO}^0
            {\textstyle\frac{1}{2}}(\bbox{\sigma}_1-\bbox{\sigma}_2)
            \!\cdot\!{\bf L} + \frac{1}{4M_{13}M_{24}}
            (\Delta\phi_C^0+\phi_C^0\Delta)\right\}. \label{Vscalr}
\end{eqnarray}

(d) Diffractive (pomeron-like) exchange:
\begin{eqnarray}
  V_D(r)=\frac{m}{4\pi}g_{13}^Dg_{24}^D \frac{4}{\sqrt{\pi}}
          \frac{m^2}{{\cal M}^2} && \left[\left\{
           1+\frac{m^2}{2M_{13}M_{24}}(3-2m^2r^2)
           +\frac{m^2}{M_{13}M_{24}}{\bf L\!\cdot\!S}
           +\frac{m^4}{4M_{13}^2M_{24}^2}Q_{12} \right. \right. \nonumber\\
  && \left. +\frac{m^2}{M_{13}M_{24}}\,
             \frac{(M_{24}^2-M_{13}^2)}{2M_{13}M_{24}}
            {\textstyle\frac{1}{2}}(\bbox{\sigma}_1-\bbox{\sigma}_2)
            \!\cdot\!{\bf L} \right\}e^{-m^2r^2}       \nonumber\\
  && \left. + \frac{1}{4M_{13}M_{24}}
            (\Delta e^{-m^2r^2}+e^{-m^2r^2}\Delta)\right]. \label{Vdifr}
\end{eqnarray}

The expressions for the configuration-space functions $\phi^n_X(r)$ can
be found in Refs.~\cite{Nag78,Mae89}, while $S_{12}$ and $Q_{12}$ are
the standard tensor and quadratic spin-orbit operators:
\begin{eqnarray}
   S_{12} &=& 3(\bbox{\sigma}_1\!\cdot\!{\bf \hat{r}})
               (\bbox{\sigma}_1\!\cdot\!{\bf \hat{r}})
              -(\bbox{\sigma}_1\!\cdot\!\bbox{\sigma}_2), \nonumber\\
   Q_{12} &=& {\textstyle\frac{1}{2}}\left[
     (\bbox{\sigma}_1\!\cdot\!{\bf L})(\bbox{\sigma}_2\!\cdot\!{\bf L})
    +(\bbox{\sigma}_2\!\cdot\!{\bf L})(\bbox{\sigma}_1\!\cdot\!{\bf L})
              \right].
\end{eqnarray}
The terms proportional to $(\Delta\phi+\phi\Delta)$ are known as the
nonlocal contributions, and represent the explicit momentum-dependent
terms (i.e., terms proportional to ${\bf q}^2$, the square of the sum
of the initial and final momenta) in the momentum-space potential.

In addition to the vector-exchange potential given in Eq.~(\ref{Vvec1r}),
there is a non-negligible contribution due to the second part of the
vector-meson propagator, $k_{\mu}k_{\nu}/m^2$. Its structure is
similar to the scalar-exchange potential given in Eq.~(\ref{Vscalr}),
and so we have
\begin{equation}
   V_V(r)\rightarrow V_V(r)-\frac{(M_3-M_1)(M_4-M_2)}{m^2}V_S(r),
                                 \label{Vvec2r}
\end{equation}
where in $V_S(r)$, obviously, now the vector-meson coupling constants
have to be used. Also, it is clear that this part only contributes when
both $M_3\neq M_1$ and $M_4\neq M_2$.

\begin{figure}
\caption{Volume integral for the scalar-exchange central $Y\!N$
         potentials in arbitrary units.}
\label{fig:scalYN}
\end{figure}

\begin{figure}
\caption{Calculated total cross sections compared with experimental data.
         Solid curve: NSC97a; dashed curve: NSC97c;
         dotted curve: NSC97f.
         Experimental data in (a) from Ref.~\protect\cite{Ale68}
         (closed circles) and Ref.~\protect\cite{Sec68} (open triangles);
         in (b) from Ref.~\protect\cite{Kad71} (closed circles)
         and Ref.~\protect\cite{Hau77} (open triangles);
         in (c) and (d) from Ref.~\protect\cite{Eis71}; 
         and in (e) and (f) from Ref.~\protect\cite{Eng66}.}
\label{fig:moddat}
\end{figure}

\begin{figure}
\caption{Predictions for differential cross sections for models NSC97a
         (solid line), NSC97c (dotted line), and NSC97f
         (dash-dotted line).
         Experimental data from Ref.~\protect\cite{Eis71}.}
\label{fig:Eisdsg}
\end{figure}

\begin{figure}
\caption{$\Lambda p$ $^1S_0$ phase shifts for models NSC97a
         (solid line), NSC97c (dotted line), and NSC97f
         (dash-dotted line).}
\label{fig:Sphs}
\end{figure}

\begin{table}
\caption{Isospin factors for the various meson exchanges in the two
         isospin channels. $P$ is the exchange operator (see text).}
\begin{tabular}{ccc}
   Matrix element & $I={\textstyle\frac{1}{2}}$
                  & $I={\textstyle\frac{3}{2}}$ \\[1mm]
\tableline
  $(\Lambda N|\eta |\Lambda N)$  &    1         &     0 \\
  $(\Lambda N|\eta'|\Lambda N)$  &    1         &     0 \\
  $(\Lambda N|\pi  |\Lambda N)$  &  $-0.0283$   &     0 \\
  $(\Lambda N|K    |N \Lambda)$  &   $P$        &     0 \\
  $(\Sigma  N|\eta |\Sigma  N)$  &    1         &     1 \\
  $(\Sigma  N|\eta'|\Sigma  N)$  &    1         &     1 \\
  $(\Sigma  N|\pi  |\Sigma  N)$  &  $-2$        &     1 \\
  $(\Sigma  N|K    |N  \Sigma)$  &  $-P$        &  $2P$ \\
  $(\Lambda N|\pi  |\Sigma  N)$  & $-\sqrt{3}$  &     0 \\
  $(\Lambda N|K    |N  \Sigma)$  & $-P\sqrt{3}$ &     0 \\
  $(\Sigma  N|\pi  |\Lambda N)$  & $-\sqrt{3}$  &     0 \\
  $(\Sigma  N|K    |N \Lambda)$  & $-P\sqrt{3}$ &     0
\end{tabular}
\label{tabisofac}
\end{table}

\begin{table}
\caption{Coupling constants, $F/(F+D)$ ratios $\alpha$, mixing angles,
         and cut-off parameters in MeV/$c^2$, common to all six models.
         Singlet refers to the {\it physical} meson, i.e., $\eta'$,
         $\omega$, $\varepsilon$, and pomeron.
         Subscripts 8, 1, and $K$ on the cut-off parameter $\Lambda$
         refer to isovector, isoscalar, and strange (isodoublet) mesons
         within the meson nonet, respectively.
         A dash means this parameter differs from one model to the next.}
\begin{tabular}{lccccdddd}
 Mesons & & Singlet & Octet & $\alpha$ & Angles
        & $\Lambda_8$ & $\Lambda_1$ & $\Lambda_K$ \\
\tableline
 Pseudoscalar & $f/\sqrt{4\pi}$ & 0.14410 & 0.27286 & 0.355
              &--23.0$^{\circ}$ & 1254.63 &  872.09 & 1281.64 \\
 Vector       & $g/\sqrt{4\pi}$ & 2.92133 & 0.83689 & 1.000
              &  37.5$^{\circ}$ &  895.07 &  949.33 & 1184.52 \\
              & $f/\sqrt{4\pi}$ & 1.18335 & 3.53174 &  --
              &                 &         &         &         \\
 Scalar       & $g/\sqrt{4\pi}$ & 4.59789 & 1.39511 &  --
              &      --         &  548.72 &  988.99 &  935.75 \\
 Diffractive  & $g/\sqrt{4\pi}$ & 2.86407 & 0.0     & 0.250
              &   0.0$^{\circ}$ &         &         &
\end{tabular}
\label{tabparcomm}
\end{table}

\begin{table}
\caption{Fitted scalar-meson mixing angle, $\theta_S$,
         and flavor-symmetry breaking parameters, $\lambda_{\rm fsb}$,
         for models NSC97a--f.
         Note that the scalar $F/(F+D)$ ratio $\alpha_S$ was not
         fitted, but is determined by Eq.~(\protect\ref{alphas}).}
\begin{tabular}{ccccccc}
 Model & $\alpha_V^m$ & $\theta_S$ & $\alpha_S$
       & $\lambda^P_{\rm fsb}$ & $\lambda^V_{\rm fsb}$
       & $\lambda^S_{\rm fsb}$ \\
\tableline
(a) & 0.4447 & 37.07$^{\circ}$ & 1.086 & 0.957 & 0.828 & 0.918 \\
(b) & 0.4247 & 37.32$^{\circ}$ & 1.091 & 1.003 & 0.895 & 0.946 \\
(c) & 0.4047 & 37.57$^{\circ}$ & 1.096 & 1.022 & 0.985 & 0.990 \\
(d) & 0.3847 & 38.31$^{\circ}$ & 1.111 & 1.084 & 1.090 & 1.037 \\
(e) & 0.3747 & 38.88$^{\circ}$ & 1.123 & 1.137 & 1.145 & 1.061 \\
(f) & 0.3647 & 39.65$^{\circ}$ & 1.138 & 1.242 & 1.188 & 1.070
\end{tabular}
\label{tabparfit}
\end{table}

\begin{table}
\caption{$\chi^2$ results on the 35 $Y\!N$ experimental total cross
         sections for the six different models, labeled according
         to the $\alpha_V^m$ input (see Table~\protect\ref{tabparfit}.
         The last column gives the predictions for the capture ratio
         at rest.}
\begin{tabular}{cccccccc}
   & $\Lambda p\rightarrow\Lambda p$ & $\Lambda p\rightarrow\Lambda p$
   & $\Sigma^+p\rightarrow\Sigma^+p$ & $\Sigma^-p\rightarrow\Sigma^-p$
   & $\Sigma^-p\rightarrow\Sigma^0n$ & $\Sigma^-p\rightarrow\Lambda n$
   & $r_{R}^{\rm th}$ \\
 Model & Ref.~\cite{Ale68} & Ref.~\cite{Sec68}
       & Ref.~\cite{Eis71} & Ref.~\cite{Eis71} & Ref.~\cite{Eng66}
       & Ref.~\cite{Eng66} & Ref.~\cite{Hep68} \\
\tableline
 (a) & 1.63 & 2.12 & 0.07 & 2.28 & 5.90 & 3.68 & 0.469 \\
 (b) & 1.59 & 2.22 & 0.06 & 2.32 & 5.82 & 3.77 & 0.466 \\
 (c) & 1.78 & 2.00 & 0.08 & 1.98 & 5.86 & 3.90 & 0.469 \\
 (d) & 1.98 & 1.93 & 0.10 & 1.89 & 5.84 & 4.01 & 0.468 \\
 (e) & 2.29 & 1.89 & 0.10 & 1.89 & 5.88 & 4.00 & 0.468 \\
 (f) & 2.52 & 2.04 & 0.20 & 1.95 & 6.01 & 3.94 & 0.467
\end{tabular}
\label{tabchimod}
\end{table}

\begin{table}
\caption{Singlet $^1S_0$ and triplet $^3S_1$ scattering lengths for
         models NSC97a--f in the different channels.}
\begin{tabular}{ccccccccc}
  & \multicolumn{2}{c}{$\Sigma^+p$} & \multicolumn{2}{c}{$\Lambda p$}
  & \multicolumn{2}{c}{$\Lambda n$} & \multicolumn{2}{c}{$\Sigma^-n$}\\
 Model & $^1S_0$ & $^3S_1$  &  $^1S_0$ & $^3S_1$
       & $^1S_0$ & $^3S_1$  &  $^1S_0$ & $^3S_1$ \\
\tableline
  (a)  & --4.35  & --0.14   & --0.71   & --2.18
       & --0.76  & --2.14   & --6.13   & --0.15  \\
  (b)  & --4.32  & --0.17   & --0.90   & --2.13
       & --0.97  & --2.08   & --6.06   & --0.18  \\
  (c)  & --4.28  & --0.25   & --1.20   & --2.08
       & --1.28  & --2.06   & --5.98   & --0.28  \\
  (d)  & --4.23  & --0.29   & --1.71   & --1.95
       & --1.82  & --1.93   & --5.89   & --0.33  \\
  (e)  & --4.23  & --0.28   & --2.10   & --1.86
       & --2.24  & --1.82   & --5.90   & --0.32  \\
  (f)  & --4.35  & --0.25   & --2.51   & --1.75
       & --2.68  & --1.66   & --6.16   & --0.29
\end{tabular}
\label{tabmodscat}
\end{table}

\begin{table}
\caption{$\Sigma^+p$ nuclear bar phase shifts in degrees for NSC97f.}
\begin{tabular}{cddddd}
 $p_{\Sigma^+}$ (MeV/$c$) &   200  & 400  &  600  &  800  &  1000  \\
  $T_{\rm lab}$ (MeV)     &  16.7  & 65.5 & 142.8 & 244.0 &  364.5 \\
\tableline
 $^1S_0$         &   42.01 &   28.67 &   11.86 &  --3.82 & --17.81 \\
 $^3P_0$         &    4.92 &    9.79 &    4.51 &  --5.48 & --16.48 \\
 $^1P_1$         &    2.55 &    9.36 &   13.70 &   12.40 &    7.39 \\
 $^3P_1$         &  --3.03 &  --9.72 & --17.07 & --24.94 & --32.92 \\
 $^3S_1$         &    7.11 &   16.10 &   28.36 &   39.79 &   43.80 \\
 $\varepsilon_1$ &  --1.90 &  --2.82 &    0.08 &    3.16 &    4.23 \\
 $^3D_1$         &    0.26 &    1.20 &    1.31 &  --1.26 &  --6.60 \\
 $^1D_2$         &    0.29 &    1.88 &    5.01 &    8.90 &   11.70 \\
 $^3D_2$         &  --0.43 &  --2.24 &  --4.23 &  --6.51 &  --9.49 \\
 $^3P_2$         &    0.79 &    4.43 &    8.01 &    9.60 &    9.83 \\
 $\varepsilon_2$ &  --0.36 &  --1.84 &  --3.00 &  --3.20 &  --2.60 \\
 $^3F_2$         &    0.03 &    0.38 &    0.83 &    0.76 &  --0.41
\end{tabular}
\label{tabphs:Sp}
\end{table}

\begin{table}
\caption{$\Lambda p$ nuclear bar phase shifts in degrees for NSC97f.}
\begin{tabular}{cddddddd}
 $p_{\Lambda}$ (MeV/$c$) & 100 & 200 & 300 & 400 & 500 & 600 & 633.4 \\
  $T_{\rm lab}$ (MeV) & 4.5 & 17.8 & 39.6 & 69.5 & 106.9 & 151.1 & 167.3 \\
\tableline
 $^1S_0$ &  25.68 &  31.52 &  28.08 &  21.52 &  14.03 &   6.42 &   3.92 \\
 $^3P_0$ &   0.02 &   0.05 & --0.39 & --2.01 & --5.10 & --9.42 &--11.00 \\
 $^1P_1$ & --0.08 & --0.59 & --1.82 & --3.88 & --6.71 &--10.08 &--11.24 \\
 $^3P_1$ & --0.09 & --0.74 & --2.38 & --5.04 & --8.47 &--12.12 &--13.06 \\
 $^3S_1$ &  19.26 &  25.92 &  24.76 &  20.57 &  15.62 &  11.55 &   7.68 \\
 $\varepsilon_1$
         &   0.16 &   0.81 &   1.80 &   3.03 &   4.77 &  10.18 &  19.81 \\
 $^3D_1$ &   0.00 &   0.05 &   0.36 &   1.49 &   5.15 &  23.26 &  76.52 \\
 $^1D_2$ &   0.00 &   0.05 &   0.30 &   0.96 &   2.08 &   3.54 &   4.07 \\
 $^3D_2$ &   0.00 &   0.08 &   0.44 &   1.27 &   2.61 &   4.32 &   4.95 \\
 $^3P_2$ &   0.05 &   0.31 &   0.59 &   0.52 & --0.16 & --1.45 & --1.99 \\
 $\varepsilon_2$
         & --0.00 & --0.01 & --0.10 & --0.31 & --0.62 & --0.99 & --1.11 \\
 $^3F_2$ &   0.00 &   0.00 &   0.01 &   0.06 &   0.19 &   0.47 &   0.70
\end{tabular}
\label{tabphs:Lp}
\end{table}

\begin{table}
\caption{$\Lambda p\rightarrow\Lambda p, \Sigma^+n, \Sigma^0p$
         total cross sections in mb above the $\Sigma N$ thresholds
         for NSC97f.}
\begin{tabular}{ddddd}
 $p_{\Lambda}$ (MeV/$c$) & $T_{\rm lab}$ (MeV)
                         & $\Lambda p\rightarrow\Lambda p$
                         & $\Lambda p\rightarrow\Sigma^+n$
                         & $\Lambda p\rightarrow\Sigma^0p$ \\
\tableline
   650  &  175.5  &  23.30  &  8.11  &  2.90 \\
   700  &  201.4  &  15.87  &  7.80  &  3.68 \\
   750  &  228.7  &  15.34  &  7.39  &  3.59 \\
   800  &  257.2  &  15.94  &  6.93  &  3.41 \\
   850  &  286.9  &  16.82  &  6.50  &  3.22 \\
   900  &  317.8  &  17.73  &  6.12  &  3.04 \\
   950  &  349.7  &  18.60  &  5.79  &  2.88 \\
  1000  &  382.6  &  19.40  &  5.50  &  2.74
\end{tabular}
\label{tabsgt:Lp}
\end{table}

\begin{table}
\caption{$\Sigma^-p\rightarrow\Sigma^-p, \Sigma^0n, \Lambda n$
         total nuclear cross sections in mb above the $\Sigma N$
         thresholds for NSC97f.}
\begin{tabular}{ddddd}
 $p_{\Sigma^-}$ (MeV/$c$) & $T_{\rm lab}$ (MeV)
                         & $\Sigma^-p\rightarrow\Sigma^-p$
                         & $\Sigma^-p\rightarrow\Sigma^0n$
                         & $\Sigma^-p\rightarrow\Lambda n$ \\
\tableline
    50  &    1.0  &  427.8  &  672.8  &  862.3 \\
   100  &    4.2  &  211.8  &  232.3  &  270.2 \\
   150  &    9.4  &  143.2  &  128.1  &  132.5 \\
   200  &   16.6  &  107.8  &   85.3  &   78.4 \\
   250  &   25.8  &   86.0  &   62.7  &   51.9 \\
   300  &   37.0  &   71.4  &   48.8  &   37.0 \\
   350  &   50.1  &   60.9  &   39.4  &   28.0 \\
   400  &   65.0  &   53.2  &   32.6  &   22.1 \\
   450  &   81.8  &   47.3  &   27.5  &   18.1 \\
   500  &  100.2  &   42.7  &   23.5  &   15.3
\end{tabular}
\label{tabsgt:Sp}
\end{table}

\begin{table}
\caption{Partial-wave contributions to the $\Lambda$ potential energy
         $U_\Lambda(k_\Lambda=0)$ at $k_F=1.35$ fm$^{-1}$
         in the cases of NSC97 models. G-matrix calculations are
         performed with the QTQ prescription for intermediate spectra.
         All entries are in MeV.}
\begin{tabular}{cddddddd}
Model & $^1S_0$ & $^3S_1$ & $^1P_1$ & $^3P_0$ & $^3P_1$ & $^3P_2$ & Sum \\
\tableline
  (a) &   --3.8 &  --30.7 & 1.5 & --0.2 & 1.6 & --2.2 & --33.9 \\
  (b) &   --5.5 &  --30.0 & 1.6 & --0.1 & 1.9 & --2.1 & --34.1 \\
  (c) &   --7.8 &  --29.7 & 1.7 &   0.2 & 2.2 & --1.9 & --35.3 \\
  (d) &  --11.0 &  --27.7 & 1.9 &   0.4 & 2.7 & --1.5 & --35.1 \\
  (e) &  --12.8 &  --26.0 & 2.1 &   0.5 & 3.2 & --1.2 & --34.3 \\
  (f) &  --14.4 &  --22.9 & 2.4 &   0.5 & 4.0 & --0.7 & --31.1
\end{tabular}
\label{tabgmat1}
\end{table}

\begin{table}
\caption{Contributions to $U_\Lambda$ at $k_F=1.35$ fm$^{-1}$ from
         spin-independent, spin-spin, $LS$, and tensor parts of the
         G-matrix interactions. See the text for the definitions of $U_0$,
         $U_{\sigma\sigma}$, $U_{LS}$, and $U_{T}$.
         All entries are in MeV.}
\begin{tabular}{cdddddd}
      & \multicolumn{2}{c}{$S$-states} & \multicolumn{4}{c}{$P$-states} \\
Model & $U_0(S)$ & $U_{\sigma\sigma}(S)$
      & $U_0(P)$ & $U_{\sigma\sigma}(P)$ & $U_{LS}(P)$ & $U_{T}(P)$ \\
\tableline
  (a) &  --8.62 & --1.61 &   0.30 & --0.39 &  --0.28 & 0.17 \\
  (b) &  --8.88 & --1.13 &   0.38 & --0.41 &  --0.32 & 0.17 \\
  (c) &  --9.37 & --0.52 &   0.46 & --0.40 &  --0.37 & 0.15 \\
  (d) &  --9.67 &   0.43 &   0.61 & --0.42 &  --0.43 & 0.15 \\
  (e) &  --9.70 &   1.04 &   0.72 & --0.44 &  --0.46 & 0.17 \\
  (f) &  --9.33 &   1.68 &   0.92 & --0.50 &  --0.47 & 0.22 \\
NSC89 &  --6.00 &   3.10 &   0.27 & --0.43 &  --0.53 & 0.14 \\
NHC-F &  --7.67 &   0.77 &   0.13 & --0.39 &  --0.49 & 0.14 \\
NHC-D &  --8.13 & --0.24 & --1.08 &   0.46 &  --0.44 & 0.09
\end{tabular}
\label{tabgmat2}
\end{table}

\begin{table}
\caption{Comparison between the QTQ and CIES treatments for the
         intermediate spectrum. See the text for the definitions
         of $U_0$, $U_{\sigma\sigma}$, $U_{LS}$, and $U_{T}$.
         All entries are in MeV.}
\begin{tabular}{ldddd}
   & \multicolumn{2}{c}{$k_F=1.35$ fm$^{-1}$}
   & \multicolumn{2}{c}{$k_F=1.0$ fm$^{-1}$}  \\
   & QTQ  & CIES & QTQ & CIES \\
\tableline
 $U_\Lambda$           & --31.1  & --34.3  & --19.9  & --21.9  \\
 $U_0(S)$              &  --9.33 &  --9.96 &  --5.25 &  --5.73 \\
 $U_{\sigma\sigma}(S)$ &    1.68 &    1.58 &    0.66 &    0.66 \\
 $U_0(P)$              &    0.92 &    0.83 &    0.18 &    0.16 \\
 $U_{\sigma\sigma}(P)$ &  --0.50 &  --0.47 &  --0.11 &  --0.10 \\
 $U_{LS}(P)$           &  --0.47 &  --0.44 &  --0.10 &  --0.09 \\
 $U_{T}(P)$            &    0.22 &    0.14 &    0.06 &    0.05
\end{tabular}
\label{tabgmat3}
\end{table}

\begin{table}
\caption{Strengths of $\Lambda$ spin-orbit splittings for various
         Nijmegen models.
         See the text for the definitions of $K_B$ and $S_{LS,ALS}$.
         The corresponding ones for the $N\!N$ interaction (G0) are
         also shown.}
\begin{tabular}{cccc}
 Model & $S_{LS}$ & $S_{ALS}$ & $K_B$  \\
\tableline
  (a)  &  --14.2 & 6.2 &  8. \\
  (b)  &  --16.2 & 6.4 & 10. \\
  (c)  &  --18.9 & 6.7 & 13. \\
  (d)  &  --21.7 & 7.1 & 15. \\
  (e)  &  --23.1 & 7.2 & 17. \\
  (f)  &  --23.9 & 7.0 & 18. \\
 NSC89 &  --28.0 & 7.9 & 21. \\
 NHC-F &  --22.8 & 5.0 & 19. \\
 NHC-D &  --22.0 & 7.3 & 15. \\
\hline
 G0($N\!N$) &  --36.4 & & 57.
\end{tabular}
\label{tabgmat4}
\end{table}

\begin{table}
\caption{Contributions to $U_\Sigma$ at $k_F=1.0$ fm$^{-1}$
         in the cases of NSC97e, NSC97f, NSC89, NHC-F, and NHC-D.
         Conversion widths $\Gamma_\Sigma$ are also shown.
         All entries are in MeV.}
\begin{tabular}{cdddddddd}
       & \multicolumn{3}{c}{Isospin $T=1/2$}
       & \multicolumn{3}{c}{Isospin $T=3/2$} & \\
 Model & $^1S_0$ & $^3S_1$ & $P$ & $^1S_0$ & $^3S_1$ & $P$
       & Sum & $\Gamma_\Sigma$ \\
\tableline
 NSC97e & 5.2 & --7.5 &   0.0 & --6.1 & --2.5 & --0.9 & --11.8 & 14.6 \\
 NSC97f & 5.2 & --7.6 &   0.0 & --6.1 & --2.2 & --0.9 & --11.5 & 15.5 \\
 NSC89  & 3.0 & --4.2 & --0.3 & --5.8 &   3.7 &   0.1 & --3.6  & 25.0 \\
 NHC-F  & 4.2 &--10.9 & --1.5 & --5.3 &  18.6 & --1.7 &   3.5  & 16.3 \\
 NHC-D  & 2.1 & --9.6 & --2.2 & --5.4 &   9.4 & --3.0 & --8.7  &  8.7
\end{tabular}
\label{tabgmat5}
\end{table}

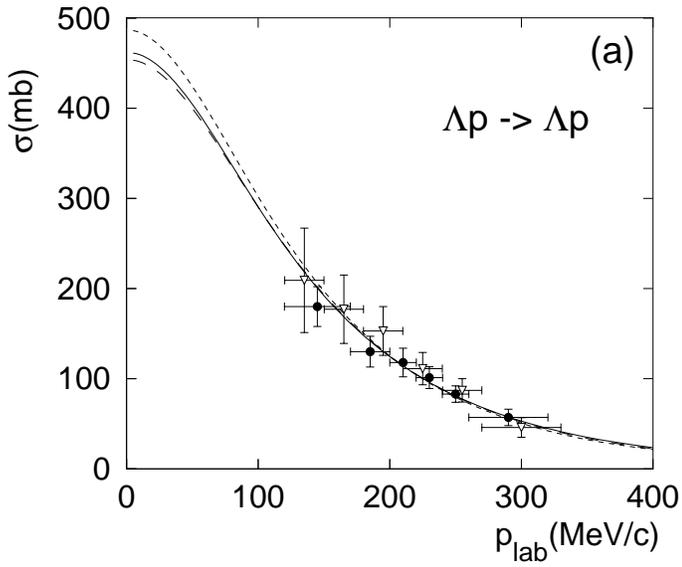
\end{document}